\documentclass[aps,prb,showpacs,preprintnumbers,twocolumn,superscriptaddress,longbibliography]{revtex4-1}
\usepackage{amsmath,amssymb}
\usepackage{bm}
\usepackage{tipa}
\usepackage{upgreek}
\usepackage{comment}
\usepackage{mathrsfs}
\usepackage{graphicx}
\usepackage{braket}
\usepackage{enumitem}
\usepackage{mathbbol}
\usepackage{booktabs}
\usepackage{gensymb}
\usepackage[normalem]{ulem}
\usepackage{color}
\usepackage[colorlinks,bookmarks=true,citecolor=blue,linkcolor=red,urlcolor=blue]{hyperref}
\usepackage{hyperref}

\usepackage{pifont}
\usepackage[dvipsnames]{xcolor}

\usepackage{float}

\newcommand{\bk}{\bm{k}}
\newcommand{\bq}{\bm{q}}
\newcommand{\bd}{\bm{d}}
\newcommand{\bG}{\bm{G}}

\begin{document}

	\title{Skyrmions in twisted bilayer graphene: stability, pairing, and crystallization} 
	\author{Yves H. Kwan}
	\affiliation{Rudolf Peierls Centre for Theoretical Physics, Parks Road, Oxford, OX1 3PU, UK}	
	\author{Glenn Wagner}
	\affiliation{Rudolf Peierls Centre for Theoretical Physics, Parks Road, Oxford, OX1 3PU, UK}
	\affiliation{Department of Physics, University of Zurich, Winterthurerstrasse 190, 8057 Zurich, Switzerland}
	\author{Nick Bultinck}
	\affiliation{Rudolf Peierls Centre for Theoretical Physics, Parks Road, Oxford, OX1 3PU, UK}
	\affiliation{Department of Physics and Astronomy, Ghent University, Krijgslaan 281, 9000 Gent, Belgium}
	\author{Steven H. Simon}
	\affiliation{Rudolf Peierls Centre for Theoretical Physics, Parks Road, Oxford, OX1 3PU, UK}
	\author{S.A. Parameswaran}
	\affiliation{Rudolf Peierls Centre for Theoretical Physics, Parks Road, Oxford, OX1 3PU, UK}

	\begin{abstract}
We study the  excitations that emerge upon doping the translationally-invariant correlated insulating states in magic angle twisted bilayer graphene at various integer filling factors $\nu$. We identify parameter regimes where  these are associated with skyrmion textures in the spin or pseudospin degrees of freedom, and explore both  short-distance pairing effects and the formation of long-range ordered skyrmion crystals.
We perform a comprehensive analysis of the  pseudospin skyrmions that emerge upon doping insulators at even  $\nu$, delineating the regime in parameter space where these are the lowest-energy charged excitations by means of  self-consistent Hartree-Fock calculations
on the interacting Bistritzer-MacDonald model. We explicitly demonstrate the purely electron-mediated pairing of skyrmions, a key ingredient behind a recent proposal of skyrmion superconductivity. Building upon this, we construct hopping models to extract the effective masses of paired skyrmions, and discuss our findings and their implications for skyrmion superconductivity in relation to experiments, focusing on the dome-shaped dependence of the transition temperature on the twist angle. We also investigate the properties of spin skyrmions about the quantized anomalous Hall insulator at $\nu=+3$. In both cases, we demonstrate the formation of robust spin/pseudospin skyrmion crystals upon doping to a finite density away from integer filling.

	\end{abstract}
	
 	\maketitle
 	
\section{Introduction}

\begin{figure*}
	\includegraphics[width=0.9\linewidth,clip=true]{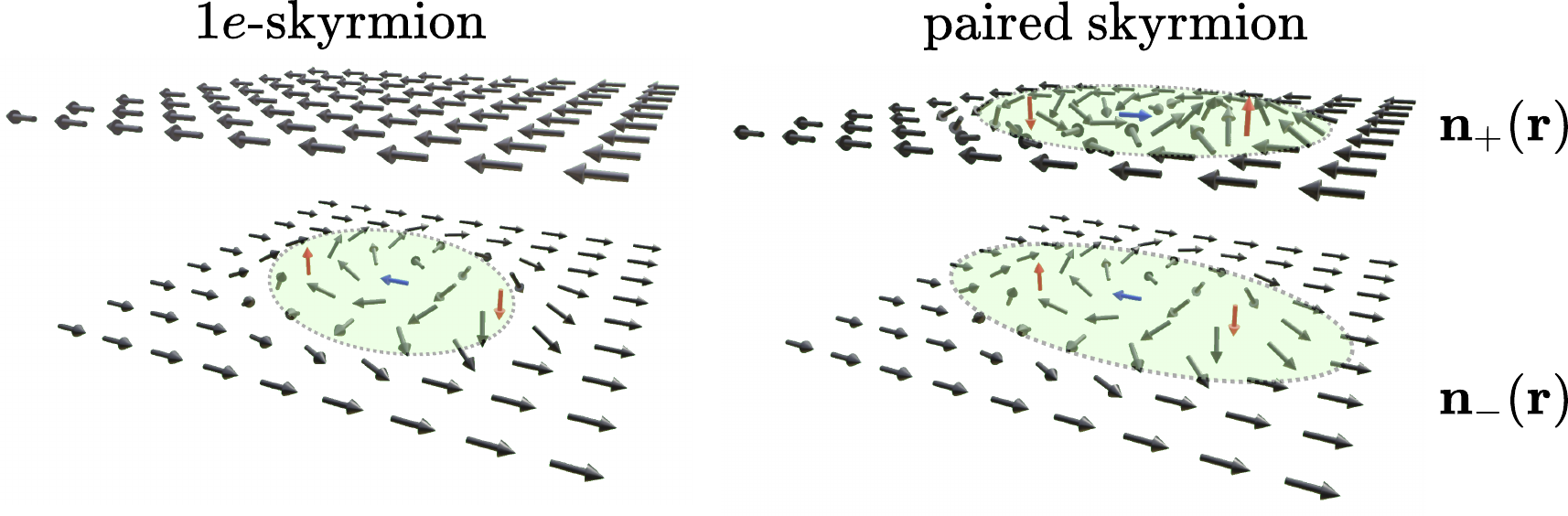}
	\caption{\textbf{Caricature of $1e$-skyrmions and paired $2e$-skyrmions at $\nu=0$.} Arrows depict the local orientation of pseudospins $\bm{n}_\pm(\bm{r})$ in the two Chern sectors ($C=\pm1$) as defined in Eq.~\ref{eq:npm} for skyrmions about the KIVC insulator at charge neutrality. Green shaded regions show where the doped electric charge is localized. Blue arrows indicate the center of the skyrmions, and red arrows denote where pseudospins are aligned along the $z$-axis.}
	\label{fig:skyrmions}
\end{figure*}

The experimental observation of an array of interaction-driven electronic phases in magic-angle twisted bilayer graphene (TBG)~\cite{Cao2018,Cao2018b,Yankowitz_2019,Lu2019}
has stimulated intense theoretical efforts to formulate a unifying physical description of  their emergence from the interplay of topology, symmetry, and correlations. One promising route to a solution has close ties to the now-classic paradigm of strongly interacting electrons in Landau levels (LLs). As shown in Ref.~\onlinecite{Bultinck_2020}, this so-called `strong-coupling' picture is rooted in the identification of an idealized limit with a $U(4)\times U(4)$ global symmetry that operates within the subspace of the eight central bands: the two $U(4)$ symmetries rotate within  quartets of topological Chern bands with Chern numbers $C=\pm1$ respectively (see also Refs.~\onlinecite{KangVafekPRL,TBG3,TBG4,vafek2020RG}). Correlated insulating phases arise as a result of symmetry breaking within this (almost) degenerate subspace, in a manner analogous to the formation of quantum Hall ferromagnetism in integrally-filled LLs~\cite{girvin1999leshouches}.  This perspective receives support from the detection of a quantized anomalous Hall insulator (QAHI) at $\nu=+3$~\cite{Sharpe_2019,Serlin900,tschirhart2021imaging} and various magnetic-field stabilized Chern insulators at other fillings~\cite{Lu2019,Saito_2021Hofstadter,Park_2021,Das_2021,stepanov2020competing,nuckolls2020strongly,Wu_2021}. While a purely strong-coupling treatment does not completely resolve the experimental puzzles surrounding TBG, it serves as a natural starting point for considering how various physically realistic corrections influence the competition between
ordered states. In this manner, it can aid in the identification of new and distinct broken-symmetry phases that are not accessible in any obvious way from the weakly-interacting limit, but can be viewed as descending from strong-coupling ferromagnetic orders in a natural, albeit non-trivial, fashion as the system is tuned towards the physically relevant intermediate-coupling regime~\cite{kwan2021kekule,wagner2021global}.

The underlying topology of TBG not only influences the selection of correlated states, but also leaves its fingerprint on the nature of their excitations
--- suggesting a tantalizing link between insulating behaviour at integer filling and superconducting behaviour away from it~\cite{khalaf2021charged,chatterjee2020skyrmionSC,Christos_2020,khalaf2020nonunitary,jing2021mechanism}. In this vein, neutral excitations of the QAHI have been predicted to inherit striking characteristics from the symmetry-breaking and topological properties of the parent insulator~\cite{wu2020collective,Kwan2021exciton,Kwan2020fqhe,Stefanidis2020}. Given the intimate connection to QHFM, the strong-coupling insulators of TBG can also be expected to host charged excitations in the form of spin or pseudospin textures known as skyrmions or merons~\cite{belavin1975metastable,sondhi1993skyrmions,moon1995,fertig1994charged,abolfath1997critical,chatterjee2020symmetry,zhang2019nearly,wu2020quantum,khalaf2021charged}. For example, under certain conditions additional charge carriers 
can enter the QAHI as skyrmions rather than as single spin-flip quasiparticles. The situation is  richer at even integer fillings since the intricate structure afforded by the approximate $U(4)\times U(4)$ symmetry and the associated anisotropies is more apparent here. Besides spin, the relevant textures now also have access to the valley and sublattice degrees of freedom (additional `flavours'), leading to the emergence of pseudospin skyrmions. The importance of a detailed understanding of such topological objects is underscored by a recent proposal~\cite{khalaf2021charged} of purely electronic superconductivity arising from bosonic `paired skyrmions'~\cite{abanov2001chiral,grover2008topological} of charge $2e$ (see Fig.~\ref{fig:skyrmions}), whose stabilization arises from the particular features of the strong-coupling Hamiltonian. Such unconventional Cooper pairs may be relevant for understanding the various superconducting domes in TBG~\cite{Cao2018b,Yankowitz_2019,Lu2019,Cao_2021,saito2020independent,Stepanov_2020,liu2021tuning,oh2021evidence}, whose properties and origins remain the subject of intense debate.

To date, however, the investigation of such flavour textures has been limited to either analyses of the effective non-linear sigma model (NLSM)~\cite{khalaf2021charged} or numerical studies of a simplified model mimicking the gross features of the central bands~\cite{chatterjee2020skyrmionSC}. While such studies provide qualitative insight,  a systematic exploration of these charged textures within a realistic microscopic model of TBG remains an outstanding challenge. There is no guarantee that  topological textures are relevant once unavoidable complications such as momentum space structure, finite kinetic energy and the presence of a moir\'e lattice are accounted for. (A recent study of domain-wall textures in the $\nu=+3$ QAHI, where microscopic energetics significantly alter properties of the topologically mandated electronic boundary modes, suggests that it is indeed important to consider such details~\cite{Kwan2020domain}.) This is the lacuna which we address in this paper, by directly probing the nature of (pseudo)spin skyrmions using unbiased self-consistent Hartree-Fock (HF) calculations within the interacting Bistritzer-MacDonald (BM) model~\cite{Bistritzer2011,santos2007}. This requires us to consider completely unrestricted Slater determinants, allowing for full breaking of moir\'e translation symmetry. This is a challenging task even within the constrained variational space of the HF mean-field ansatz, but one that we pursue successfully below.

We have three main objectives in this work: first, to provide a proof-of-principle demonstration of  skyrmion pairing  in an appropriately-chosen limit of a microscopic model of TBG; second, to investigate whether such pairing can result in superconducting phenomenology consistent with that seen in experiment (most notably, the dome-shaped dependence of the superconducting transition temperature $T_c$ on the twist angle~\cite{Cao_2021}); third, to establish the regime in parameter space over which  pairing survives away from the idealized limit. We also consider a subsidiary set of issues pertaining to the emergence of skyrmions at other fillings, and their formation of long-range ordered structures~\cite{brey1995skyrme,green1996skyrmions,bomerich2020skyrmion} when doped into TBG at a finite density. 

Within a HF treatment of the interacting BM model, we find that pseudospin skyrmion pairing does indeed occur about even integer filling and at sufficiently small chiral ratio, and in qualitative agreement with the predictions of the NLSM. Paired skyrmions are lower in energy than particle excitations for small chiral ratio, with relatively better energetics at charge neutrality than at $|\nu|=2$. For chiral ratios approaching the realistic values, they become energetically unfavorable and harder to stabilize within mean-field theory. External perturbations such as strain and substrate, as well as increased screening of interactions, tend to also have the same effect. We devise a effective skyrmion hopping model to estimate the effective masses, and find that the corresponding superconducting critical temperature $T_c$~\cite{nelson1977universal} can be larger than typical scales observed experimentally. However, more detailed comparison of the twist-angle dependence is complicated by the sensitivity of the results to how electron interactions are incorporated (i.e.~the `subtraction scheme', to be explained in Sec.~\ref{subsec:IBM}). We critically discuss our findings in the context of experiments, and conclude that, at least for some of the samples, skyrmion pairing is unlikely to be mechanism for superconductivity. For the QAHI at $|\nu|=3$, we demonstrate the formation of spin skyrmions for doping of single charges, and various spin texture lattices at larger dopings.

The balance of this paper is organized as follows. After reviewing the basics of the NLSM and strong-coupling picture of TBG in Sec.~\ref{sec:theoretical}, we focus on the pseudospin skyrmions about the spinless insulator at charge neutrality ($\nu=0$) in Sec.~\ref{sec:even_integer}. Via an effective skyrmion hopping model, we further estimate the dispersion and effective mass of delocalized skyrmions. We also briefly address the situation at $|\nu|=2$ in the spinful case. In Sec.~\ref{sec:spin}, we discuss the properties of spin skyrmions and spin texture lattices at $|\nu|=3$ about the QAHI. We end with a discussion in Sec.~\ref{sec:discussion}.

\section{Theoretical considerations}\label{sec:theoretical}

In this section, we review the key concepts that underpin the existence and properties of skyrmions about the strong-coupling insulators in TBG~\cite{Bultinck_2020,TBG4,ledwith2021pedagogical}.

\subsection{Interacting BM model}\label{subsec:IBM}

\begin{figure*}
	\includegraphics[width=0.75\linewidth,clip=true]{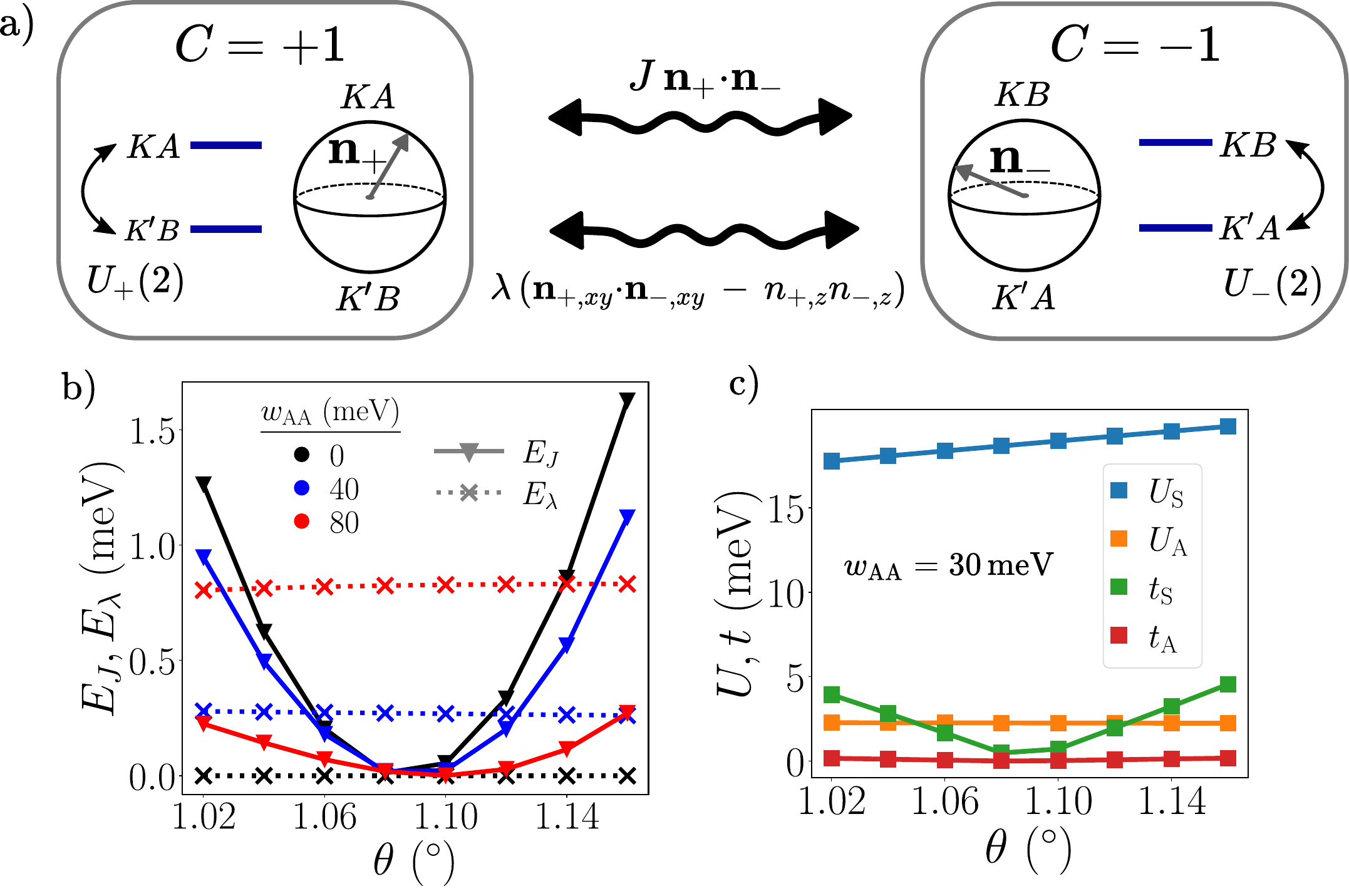}
	\caption{\textbf{Strong-coupling hierarchy and Chern pseudospins.} a) Schematic of Chern sectors with $C=\pm1$. In the isotropic limit, there is a $U(2)\times U(2)$ symmetry ($U(4)\times U(4)$ with spin). At charge neutrality for a generalized ferromagnet with net $C=0$, one can define pseudospins $\bm{n}_{\pm}$. The dominant corrections can be captured as anisotropies $J,\lambda$ (wiggly lines) which couple the Chern sectors and partially lift the symmetry. 
    b) NLSM parameters $E_J=JA_{\text{UC}}$ (triangles), $E_\lambda=\lambda A_\text{UC}$ (crosses) as a function of twist angle $\theta$ and interlayer hopping parameter $w_{{AA}}$, where $A_\text{UC}$ is the moir\'e unit cell area. The pseudospin stiffness $\rho_\text{ps}\simeq 1.5\,\text{meV}$, calculated in the isotropic limit, does not depend on $w_\text{AA}$ and subtraction scheme (see  App.~\ref{app:BM_HF}), and is largely insensitive to $\theta$. c) Strong-coupling energy scales $U_\text{S},U_\text{A},t_\text{S},t_\text{A}$ as a function of $\theta$ and $w_{{AA}}$. All quantities calculated using the average subtraction scheme. 
	\label{fig:Jlambrhos}}
\end{figure*}

The starting point is the continuum BM model $\hat{H}^{\text{BM}}$ which generates the single-particle moir\'e bands of TBG in the moir\'e Brillouin zone (mBZ)~\cite{Bistritzer2011}. Details of the model are relegated to App.~\ref{app:BM_HF}, but we mention here the important parameters which will influence the skyrmions.  The interlayer coupling is parameterized through two sublattice-dependent hopping constants $w_{\text{AA}}$ and $w_{\text{AB}}$, reflecting corrugation effects in the superlattice. We fix $w_{\text{AB}}=110\,\text{meV}$ but consider variable $w_{\text{AA}}$, thereby tuning the value of the chiral ratio $\kappa\equiv w_{\text{AA}}/w_{\text{AB}}$. Estimated values for $\kappa$ in the literature range from $0.5$ to $0.8$\cite{ledwith2021TB,Nam2017,Carr2019}. The kinetic bandwidth is minimal when the twist angle $\theta$ is close to the magic angle, which is weakly dependent on $\kappa$. While the physical relevance of $\theta$ as a tuning parameter is obvious, it will prove theoretically useful to vary $w_\text{AA}$ to control deviations from the fully symmetric theory, to be described below. (Note that it has  been argued that the effective value of $w_\text{AA}$ may be susceptible to renormalization towards the chiral limit~\cite{Tarnopolsky2019}, $\kappa=0$, due to the effects of remote bands; thus, tuning $w_\text{AA}$  may be viewed as phenomenologically modeling this downward renormalization~\cite{vafek2020RG}.)

Sometimes, we will also incorporate heterostrain~\cite{Bi2019,parker2020straininduced,kwan2021kekule} with strength $\epsilon$ and a sublattice potential of strength $\Delta$~\cite{jung2015origin}. In the absence of such single-particle perturbations, the point group is $D_6$, which includes $\hat{C}_{2z},\hat{C}_{3z}$ and $\hat{M}$ (in-plane rotation about the $x$-axis). Accounting for both valley and spin degrees of freedom, the BM model contains a $U(2)_K\times U(2)_{K'}$ flavour symmetry which includes charge/valley conservation and independent $SU(2)$ spin-rotations within each valley. Naturally, a spinless time-reversal symmetry (TRS) $\hat{\mathcal{T}}=\tau_x\mathcal{\hat{K}}$ is also present, where $\mathcal{\hat{K}}$ is the complex conjugation operator.
Upon neglecting the weak relative rotation of the Dirac cones in the two layers,
there is also a particle-hole symmetry (PHS) $\hat{\mathcal{P}}$~\cite{Zou2018approximants,song2019,hejazi2021}. Throughout this work we will focus solely on the central eight bands, and project out the remote bands which are separated by a large band gap. (Although, see the preceding comments on their possible role in adjusting $w_\text{AA}$.)

Repulsive interactions are implemented by augmenting the BM model with dual-gate screened density-density Coulomb interactions $V(q)=\frac{e^2}{2\epsilon_0\epsilon_r q}\tanh qd_\text{sc}$, where the relative permittivity $\epsilon_r=10$ and the screening length $d_\text{sc}=25$\,nm~\cite{Bultinck_2020}. By neglecting subleading `intervalley-Hund's couplings' which scatter electrons between valleys, we retain the global $U(2)_K\times U(2)_{K'}$ symmetry. 

Note that interactions may alter the one-body terms somewhat, leading to a total effective single-particle dispersion $\hat{H}^{\text{SP}}$~\cite{XieSub,Liu2021nematic,Bultinck_2020,TBG3,TBG4,hejazi2021}. The precise correction depends on the choice of `subtraction scheme' which arises as follows (a more detailed exposition can be found in App.~\ref{app:BM_HF}). Including only  normal-ordered interactions within the central band subspace neglects contributions from the remote bands. Furthermore, some interactions are double-counted because the hopping parameters of 
graphene (e.g.~from density functional theory calculations) are obtained self-consistently with filled valence bands. A convenient parameterization to remedy this and restore PHS is in terms of a reference density matrix $P^0$. There is no unique prescription for fixing $P^0$ and we will consider three choices that have been used in the literature: (i) the `average', (ii) `charge neutrality' (CN), and (iii) `graphene' schemes. 
This nomenclature is motivated by how $P^0$ is constructed: (i) $P^0\propto I$, (ii) $P^0$ consists of the filled valence BM bands, and (iii) $P^0$ consists of the filled valence bands of decoupled layers of graphene.   
Most of the initial discussion on skyrmions will be based on the average scheme, which leads to no corrections when interactions are recast in the strong-coupling form that highlights the strong-coupling hierarchy of scales (see Sec.~\ref{subsec:strong_coupling} and App.~\ref{app:BM_HF}). The other two schemes will be analyzed in more detail in Sec.~\ref{subsec:effective_mass}.

\subsection{Strong-coupling hierarchy}\label{subsec:strong_coupling}

Since the typical Coulomb scale exceeds the kinetic bandwidth, the BM band basis may not be the most natural one in which to understand interaction physics. A more appropriate choice, that reveals the underlying topological character of the bands, is furnished by the Chern basis obtained by diagonalizing the sublattice operator $\sigma_z$~\cite{Bultinck_2020}. Each band is predominantly polarized on one sublattice, which allows us to label the central bands with sublattice $\sigma$, valley $\tau$, and spin $s$. Combining these into a single index $\alpha$, we can write the Bloch functions as $\ket{\psi_\alpha(\bm{k})}=e^{i\bm{k}\cdot\hat{\bm{r}}}\ket{u_{\alpha}(\bm{k})}$ with associated creation operators $\hat{d}^\dagger_{\alpha}(\bm{k})$. Crucially, these bands have Chern number $C=\sigma_z\tau_z$ (Fig.~\ref{fig:Jlambrhos}a). 

The link to QHFM is sharpened in the well-studied chiral limit $\kappa=0$~\cite{Tarnopolsky2019,Ledwith2020FCI}, where the kinetic bandwidth becomes exactly zero at the magic angle and the Chern bands become completely sublattice polarized. In addition, the new chiral symmetry $\{\sigma_z,H^{\text{BM}}\}$, in concert with $\hat{C}_{2z}\hat{\mathcal{T}}$ and $\hat{\mathcal{P}}$, heavily constrains the form factors $\Lambda_{\alpha,\beta}(\bm{k},\bm{q})\equiv \braket{u_\alpha(\bm{k})|u_\beta(\bm{k}+\bm{q})}$ to be diagonal and to depend only the Chern number: this allows us to write $\Lambda_{\alpha,\beta}(\bm{k},\bm{q}) =  \Lambda^{\text{S}}(\bm{k},\bm{q})\delta_{\alpha\beta}$, with\cite{Bultinck_2020}
\begin{equation}
    \Lambda^{\text{S}}(\bm{k},\bm{q})=F^{\text{S}}(\bm{k},\bm{q})e^{i\phi^{\text{S}}(\bm{k},\bm{q})\sigma_z\tau_z},
\end{equation}
where the superscript `S' denotes `symmetric'. In the chiral-flat
limit where the effective dispersion $\hat{H}^{\text{SP}}$ is neglected, this implies a huge $U(4)_{C=1}\times U(4)_{C=-1}$ symmetry, as can be seen by checking the invariance of the density operator $\hat{\rho}^{\text{S}}(\bm{q})=\sum_{\bm{k}}\hat{\bm{d}}^\dagger(\bm{k})\Lambda^{\text{S}}(\bm{k},\bm{q})\hat{\bm{d}}(\bm{k}+\bm{q})$ under independent Chern-number preserving rotations. Remarkably, at $\nu=0$ one can prove the existence of exact Slater determinant ground states, which simply involve uniformly polarizing any four orthogonal directions within the Chern quartets (`ferromagnets')~\cite{Bultinck_2020,TBG4}. Hence, the chiral-flat limit will also be referred to as the isotropic limit.

The utility of the chiral-flat limit lies in the existence of a hierarchy of scales which permits corrections to be treated perturbatively within the manifold of strong-coupling ferromagnets~\cite{Bultinck_2020,TBG4}. The energy scale of the full symmetry group is $U_\text{S}$, and the main competing scales are the single-particle inter-Chern tunneling ($t_\text{S}$) and deviation from chirality ($U_\text{A}$), with intra-Chern dispersion ($t_A$) being subleading [see Fig.~\ref{fig:Jlambrhos}d and App.~\ref{app:BM_HF}]. Focusing for simplicity on the spinless insulator at neutrality such that the parent symmetry group is $U(2)\times U(2)$, the effects of the above perturbations can be captured via anisotropies with positive strengths $J$ and $\lambda$ respectively (i.e. energy scales $E_J=A_\text{UC}J$ and $E_\lambda=A_\text{UC}\lambda$ where $A_\text{UC}$ is the moir\'e unit cell area). It turns out that the states that get energetically selected by these terms have total $C=0$. Defining an intra-Chern Pauli triplet $\bm{\eta}=(\sigma_x\tau_x,\sigma_x\tau_y,\tau_z)$, we can parameterize the strong-coupling ferromagnets residing in the Chern-neutral sector with two Chern-filtered pseudospins  
\begin{equation}\label{eq:npm}
    \bm{n}_\pm(\bm{k})\equiv  \bra{\Psi}\hat{\bm{d}}^\dagger(\bm{k}) \bm{\eta} \frac{1\pm\sigma_z\tau_z}{2}  \hat{\bm{d}}(\bm{k})\ket{\Psi}
\end{equation}
which are independent of $\bm{k}$ (Fig.~\ref{fig:Jlambrhos}a). Inter-Chern tunneling generates superexchange, inducing an anti-ferromagnetic (AFM) coupling between the two pseudospins, while a finite $\kappa$ manifests as a AFM coupling in-plane, and an FM coupling out-of-plane. Both break the chiral-flat symmetry to distinct $U(2)$ subgroups. The resulting ground state with $\bm{n}_+=-\bm{n}_-$ (and hence valley-$U(1)_V$ degeneracy) is the Kramers intervalley-coherent insulator (KIVC), so called because it preserves a modified TRS $\hat{\mathcal{T}}'=\tau_y\hat{\mathcal{K}}$. An equivalent description of the spinless KIVC is via its density matrix $P=\frac{1}{2}(1+Q)$, where
\begin{equation}\label{eq:QKIVCspinless}
    Q_{\text{KIVC}}=\sigma_y(\tau_x\cos\phi_\text{IVC}+\tau_y\sin\phi_\text{IVC})
\end{equation}
is parametrized through the IVC angle $\phi_\text{IVC}$.

\subsection{Non-linear sigma model and pseudospin textures}\label{subsec:NLSM}

To understand low-energy excitations and spatially non-uniform configurations, it is useful to consider a continuum NLSM description purely in terms of the pseudospins~\cite{khalaf2021charged}. The energy functional, familiar from QHFM, is
\begin{align}\label{eq:NLSM}
\begin{split}
    E[\bm{n}_+,\bm{n}_-]=&\int d^2\bm{r}\,\bigg[\frac{\rho_{\text{ps}}}{2}\left((\nabla \bm{n}_+)^2+(\nabla\bm{n}_-)^2\right)\\
        &+(J+\lambda)\bm{n}_+\cdot\bm{n}_--2\lambda n^z_+n^z_-\\
        &+\frac{1}{2}\int d^2\bm{r}'\,\delta\rho(\bm{r})V(\bm{r}-\bm{r}')\delta\rho(\bm{r}')
\end{split}
\end{align}
where $\rho_{\text{ps}}$ is the pseudospin stiffness in the isotropic limit, and $\delta\rho=\delta\rho_++\delta\rho_-$ consists of the topological charge (Pontryagin index) densities of the two Chern sectors
\begin{equation}
    \delta\rho_\pm(\bm{r})=\pm e \rho_{\text{top},\pm}(\bm{r})=\pm\frac{ e}{4\pi}\bm{n}_\pm\cdot\partial_x\bm{n}_\pm\times\partial_y\bm{n}_\pm.
\end{equation}
The skyrmion number is given by
\begin{equation}
    N_{\text{top},\pm}=\int d^2\bm{r}\,\rho_{\text{top},\pm}(\bm{r}),
\end{equation}
where skyrmions and antiskyrmions  have $N_{\text{top}} = +1$ and $N_{\text{top}} = -1$ respectively.

The NLSM parameters $J,\lambda,\rho_{\text{ps}}$ are plotted in Fig.~\ref{fig:Jlambrhos}, with explicit expressions  supplied in App.~\ref{app:BM_HF}. In the ``average scheme" described there, the superexchange $J$ has a prominent minimum and nearly vanishes around $\theta\simeq1.08^\circ$, coincident with where the bare BM bandwidth is smallest. $\lambda$ is a monotonically increasing function of $w_\text{AA}$ and is largely insensitive to $\theta$. The pseudospin stiffness $\simeq 1.3\,\text{meV}$ is a property of the maximally-symmetric manifold and has a very weak dependence on twist angle. While there are sizable regions where $\lambda>J$, this only occurs for the average scheme where $J$ is directly tied to the bare BM scale, which is substantially smaller than the bandwidth obtained via STM measurements of the van Hove singularities in the density of states
~\cite{Kerelsky2019,Choi2019,jiang2019charge,Xie2019stm,Wong_2020}. 
Therefore we will mostly be interested in the case $J\gtrsim \lambda$, that is conducive to skyrmion pairing as explained below.

Consider first the isotropic limit $J=\lambda=0$ where the Chern sectors decouple. The ground state at $\nu=0$ is specified by free choices of uniform pseudospins $\bm{n}^0_\pm$. An additional charge enters as a 1$e$-skyrmion (in one of the Chern sectors) rather than a particle-like excitation if the energy for a well separated skyrmion-antiskrymion pair $(8\pi\rho_{\text{ps}})$ exceeds the particle-hole gap $(\Delta_{\text{ph}})$. We can use the infinite-size Polyakov solution~\cite{belavin1975metastable} which minimizes the gradient energy because of the lack of any Zeeman-like term, meaning that the texture prefers to expand to minimize the Coulomb term.

Reintroducing dispersion leads to an effective AFM coupling $J$, constraining the parent ground state $\bm{n}^0_+=-\bm{n}^0_-$ which now belongs to an $SO(3)$ manifold. A skyrmion of radius $R_\text{s}$ in say $\bm{n}_+$ now experiences a Zeeman penalty $\propto R_\text{s}^2$ arising from misalignment with $\bm{n}^0_-$. The skyrmion becomes finite with a size determined from the competition between $J$ and the interaction term. (Note that in order to  avoid divergences with system size, the tails of the skyrmion profile must decay faster in space than the Polyakov ansatz~\cite{sondhi1993skyrmions,lejnell1999effective,khalaf2021charged}.) For large enough $J$, the skyrmion shrinks and crosses over to an ordinary particle-like excitation. Having instead a finite $\lambda$ leads to similar conclusions, except with a different residual $SU(2)$ manifold. Including both perturbations restricts the insulator to a KIVC parameterized by a $U(1)$ valley angle. 

Adding instead a net charge of $|Q|=2e$ to the system leads to substantially different conclusions. In the scenario with non-zero $J$, the AFM interaction  leads to the pairing of a skyrmion in $\bm{n}_+$ with an antiskyrmion in $\bm{n}_-$ (the total charge is $2e$, a crucial consequence  of the opposite assignment of Chern numbers). To see this, note that the AFM inter-Chern coupling is completely satisfied if the two textures are centered at the same position with exactly the same profile such that $\bm{n}_\text{pair}(\bm{r})\equiv \bm{n}_+(\bm{r})=-\bm{n}_-(\bm{r})$. In this case the resulting `paired $2e$-skyrmion' dilates without limit to avoid the Coulomb self-energy. Hence an $\bm{n}_+$-skyrmion and an $\bm{n}_-$-antiskyrmion will bind for any positive value of $J$, even though the underlying electron-interaction is purely repulsive. A paired skyrmion configuration preserves $\hat{\mathcal{T}}'$-symmetry, which is a useful numerical diagnostic.

In the presence of an additional $\lambda$ term, the paired skyrmion experiences an energy penalty $\propto R_\text{s}^2$ from regions where $\bm{n}_\text{pair}(\bm{r})$ is not lying in-plane. This not only leads to a finite size, but also elongates the texture somewhat to reduce the area spent pointing out-of-plane.  When $\lambda$ is comparable to $\rho_{\text{ps}}$, the texture deforms to resemble the topologically equivalent configuration of two paired $1e$-merons separated by a finite distance. This can be understood as the $\lambda$-term shrinking the costly $\bm{n}_{\text{pair}}\parallel \hat{z}$ regions such that they become the cores of the merons (see red arrows in Fig.~\ref{fig:skyrmions}).
If $\lambda/J$ is sufficiently large, pairing is no longer favorable.

\subsection{Skyrmion superconductvity}

We now assume that the microscopic parameters are chosen such that charges enter as skyrmions rather than particles. $1e$-skyrmions from opposite Chern sectors attract to form paired skyrmions. Even if particle excitations are slightly lower in energy, this typically occurs in a small region of the mBZ centered at $\Gamma_{\text{M}}$. Hence above a critical doping, additional charges are expected to form skyrmions~\cite{khalaf2021charged}.  

A non-zero superconducting $T_c$ requires a finite boson effective mass $M_\text{pair}$, which can be motivated within the phenomenological picture of a pair of coupled quantum Hall ferromagnets~\cite{khalaf2021charged} in equal and oppposite magnetic fields. A single $1e$-skyrmion feels a net magnetic field and hence has a flat-band dispersion. On the other hand, the magnetic fields experienced by a paired skyrmion cancel out. To understand the generation of an effective mass in this case, imagine a paired skyrmion propagating at some velocity $v$. The Lorentz forces in the two sectors act to push the constituent $1e$-skyrmions in opposite directions, which is counteracted by a restoring force depending linearly on $J$ and the separation ($\lambda$ is assumed small). By balancing the two, one can deduce that $M_\text{pair}$ is inversely proportional to $J$. Therefore the paired skyrmions condense with a finite superfluid stiffness and associated Berezenskii-Kosterlitz-Thouless transition temperature~\cite{nelson1977universal}  
\begin{equation}\label{SF_Tc}
    T_c=\frac{\nu\pi\hbar^2}{2k_BA_\text{UC}M_\text{pair}}=\frac{\nu J A_\text{UC}}{2k_B}.
\end{equation}
We define a corresponding filling-independent energy scale $E_c=k_BT_c/\nu$, which equals $E_J/2$ in this framework. Above, $\nu$ should be taken to be the doping relative to the integer filling of interest.

Note the sensitive dependence on the superexchange $J$, which potentially provides a litmus test for the applicability of this mechanism to realistic samples of TBG. Namely, experiments observe a superconducting dome in $T_c$ as a function of twist angle, which peaks around $\theta=1.08^\circ$~\cite{Cao_2021}. The value of $J$, and indeed the effective mass as computed independently in Sec.~\ref{subsec:effective_mass}, are also affected by $\theta$ through the evolution of the bandwidth and character of the BM bands. However as we discuss later, the precise relationship is subtle and differs based on the choice of interaction. In particular, while the nature of the ordered normal state at $\nu=0$ is unambiguous, depending only on the signs of $J$ and $\lambda$, finer details such as skyrmion effective masses are susceptible to the specifics of the subtraction scheme (App.~\ref{app:BM_HF}).

\begin{figure*}
	\includegraphics[ width=0.8\linewidth,clip=true]{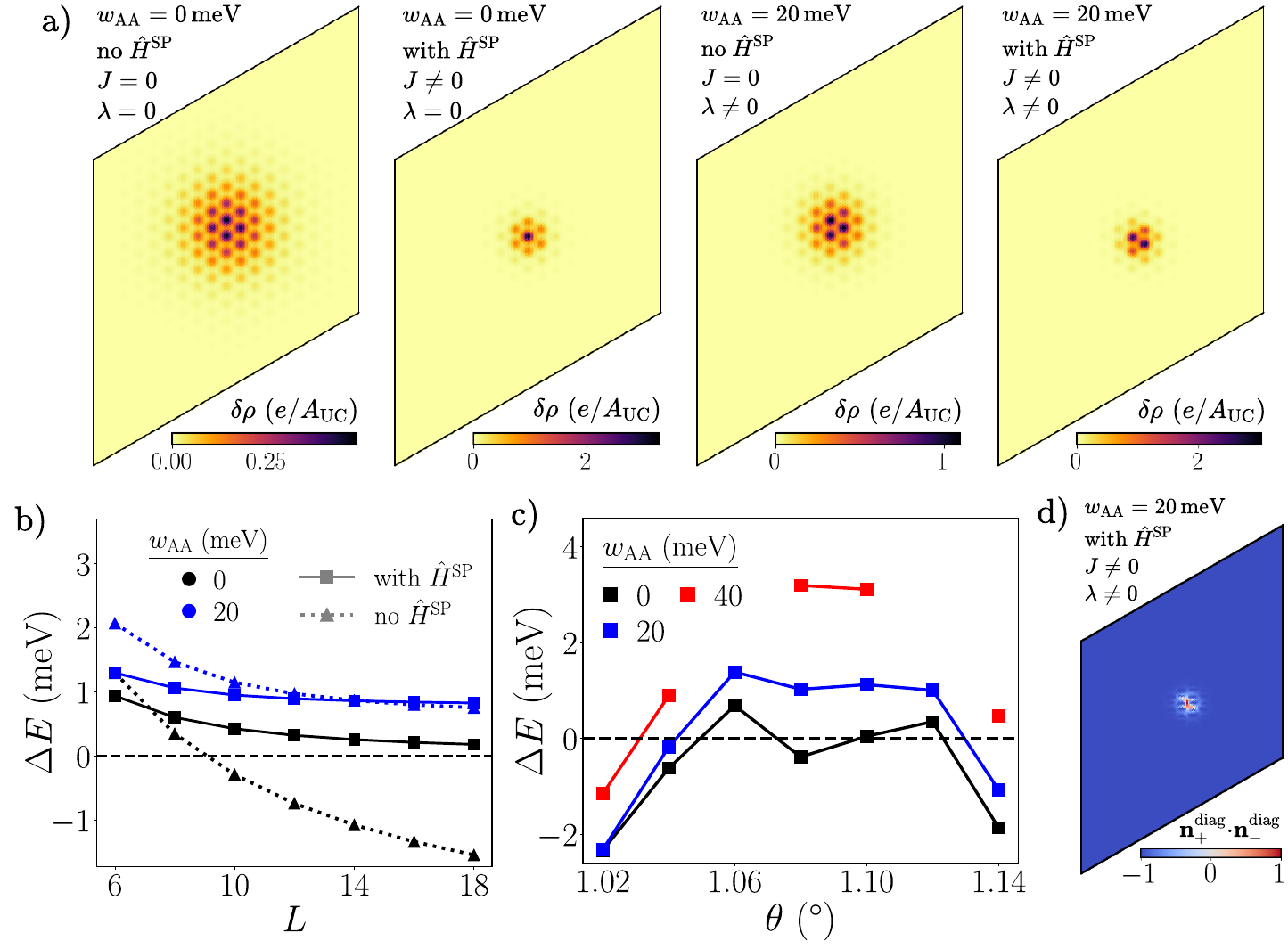}
	\caption{\textbf{Properties of $1e$-skyrmions (spinless $\nu=0$).} a) Charge density $\delta \rho(\bm{r})$ of $1e$-skyrmions in the entire simulation cell relative to $\nu=0$ ground state. Panels show different combinations of turning on/off dispersion and achirality, which determine $J$ and $\lambda$. Calculations performed on $18\times18$ systems at $\theta=1.05^\circ$. Textures manually translated to the center for clarity. b) Relative energy $\Delta E=E_{1e\text{-skyr}}-E_{1e}$ of $1e$-skyrmions (compared to particle excitations) as a function of system size. Apart from the isotropic limit (black triangles), $1e$-skyrmions are not favored.
	c) $\Delta E$ for different $\theta$ and $w_{AA}$. Dispersion is always included. System size is $11\times11$. Missing data points indicate parameters where HF was unable to find skyrmion solutions. d) Local alignment of Chern pseudospins corresponding to the rightmost panel of a). All calculations used the average scheme (see  App.~\ref{app:BM_HF}). 
    }
	\label{fig:1eskyrmions}
\end{figure*}

\subsection{Numerical methods}

Underpinning the discussion in the following sections are explicit mean-field solutions $\ket{\phi}$ for the (pseudo)spin skyrmions, constructed using unbiased self-consistent HF calculations. Since such configurations break translation symmetry and flavor rotation symmetries, the HF density matrix takes the most general form
\begin{equation}
    P_{\bm{k}\tau s \sigma;\bm{k}'\tau' s'\sigma'}=\langle \hat{d}^\dagger_{\bm{k}\tau s \sigma}\hat{d}^{\phantom{\dagger}}_{\bm{k}'\tau' s' \sigma'} \rangle.
\end{equation}
The mBZ momenta $\bm{k}$ take discrete values due to the finite system dimensions $L_1,L_2$ measured in units of the moir\'e lattice vectors $\bm{a}^\text{M}_{1},\bm{a}^\text{M}_{2}$. Throughout this work we take $L=L_1=L_2$ so that a non-degenerate moir\'e band can accommodate $L^2$ electrons. The filling is fixed by the constraint $\text{Tr}P=(4+\nu) L^2=(4+\nu_0) L^2+N$, where $\nu_0$ is the parent integer filling factor and $N$ represents the number of additional doped electrons. 
We typically initialize the calculations with random initial projectors, and accelerate convergence using the optimal damping algorithm~\cite{ODA}. Near $\nu=0$, we neglect spin for simplicity and hence omit the index $s$ when calculating both skyrmion and particle excitations.

For spin textures about the QAH state, the local order parameter is simply the local spin density
\begin{equation}
    \bm{S}(\bm{r})=\sum_{ss'\tau I}\langle \hat{\psi}^\dagger_{\tau s I}(\bm{r})\bm{s}_{ss'}\hat{\psi}^{\phantom{\dagger}}_{\tau s' I}(\bm{r}) \rangle
\end{equation}
where $\bm{s}$ is a Pauli triplet in spin space, and $I=\{1A,1B,2A,2B\}$ labels the types of electrons living on different layers/sublattices. Replacing $\bm{s}$ by the identity gives the local charge density. The real-space fermion creation operator is defined in terms of the Chern basis as
\begin{equation}
    \hat{\psi}^\dagger_{\tau s I}(\bm{r})=\sum_{\bm{k}\sigma}\braket{\psi_{\tau s \sigma}(\bm{k})|\bm{r};I}\hat{d}^\dagger_{\tau s \sigma}(\bm{k}).
\end{equation}

For pseudospin textures, the natural extension 
is the `diagonal' Chern-filtered pseudospin density
\begin{equation}\label{eq:diagdensity}
    \bm{n}^{\text{diag}}_C(\bm{r})=\sum_{I}\langle \hat{\bm\psi}^\dagger_{I}(\bm{r})\frac{1+C\tau_z\sigma_z}{2}\bm{\eta}\hat{\bm\psi}^{\phantom{\dagger}}_{I}(\bm{r}) \rangle
\end{equation}
where spin has been dropped and $\bm{\eta}$ is the triplet of pseudospinor Paulis defined above Eq.~\ref{eq:npm}. The net global Chern polarization is defined by the difference in fractional occupation in each Chern sector. Above, the Chern band-filtered real-space creation operator is 
\begin{equation}
    \left[\hat{\psi}^\dagger_{ I}(\bm{r})\right]_{\tau\sigma}=\sum_{\bk}\braket{\psi_{\tau\sigma}(\bm{k})|\bm{r};I}\hat{d}^\dagger_{\tau\sigma}(\bm{k}).
\end{equation}
where $\tau\sigma$ labels the Chern band. However, the in-plane components of Eq.~\ref{eq:diagdensity} identically vanish in the chiral limit $\kappa=0$, where the Chern basis becomes completely sublattice polarized. For example, for the $C=+1$ Chern sector, $n_{+,x}$ and $n_{+,y}$ should measure coherence between the $KA$ and $K' B$ bands, but $[\hat{\psi}^\dagger_{I}]_{KA}$ $([\hat{\psi}^\dagger_{I}]_{K'B})$ is non-zero only when $I=1A,2A$ $(1B,2B)$. 

To remedy this, we occasionally consider an alternative definition that includes off-diagonal terms in the layer/sublattice degrees of freedom
\begin{equation}
    \bm{n}^{\text{off}}_C(\bm{r})=\sum_{I I'}\langle \hat{\bm\psi}^\dagger_{I}(\bm{r})\frac{1+C\tau_z\sigma_z}{2}\bm{\eta}\hat{\bm\psi}^{\phantom{\dagger}}_{I'}(\bm{r}) \rangle.
\end{equation}

To address the question of the effective mass, we can perform variational calculations in the space of states obtained by translating a localized paired skyrmion $\ket{\phi}$ by all possible moir\'e lattice vectors. This defines an effective skyrmion hopping model leading to delocalized `Bloch skyrmions' and a Bloch dispersion, from which $M_\text{pair}$ can be extracted. Note that this way of calculating the paired skyrmion mass is completely different from the classical calculation of $M_{pair}$ used to obtain the second equality of Eq.~\ref{SF_Tc}. The technical details of the skyrmion-plane wave calculations, including generalizations involving symmetry-related skyrmions, are outlined in App.~\ref{app:effective}.

\section{Pseudospin textures at even integer filling}\label{sec:even_integer}

\subsection{$1e$-skyrmions}

Fig.~\ref{fig:1eskyrmions}a shows the charge density of a 1e-skyrmion relative to the translation-invariant $\nu=0$ ground state. The Chern couplings $J$ and $\lambda$ can be controlled by tuning the effective one-body term and the chiral ratio respectively. Since the Hamiltonian obeys particle-hole symmetry, adding holes instead of electrons leads to analogous results. 

In the isotropic limit with $J=\lambda=0$, the ground state can be any member of the $U(2)\times U(2)$ manifold described by independent choices of $\bm{n}_+,\bm{n}_-$. (Strictly speaking the $|C|=2$ ferromagnets are also ground states, but these do not allow for textures). As verified numerically, the extra charge enters as a delocalized skyrmion solely in one of the Chern sectors, while the other sector remains unchanged. Hence its properties are qualitatively the same as for spin skyrmions about the $\nu=3$ QAH insulator (see Sec.~\ref{sec:spin}). Consistent with the absence of anisotropies, the skyrmion expands as much as possible and is only limited by the finite simulation cell. This explains the slow convergence in Fig.~\ref{fig:1eskyrmions}b of the $1e$-skyrmion energy (measured with respect to the $1e$-particle excitation) with the linear system dimension $L$, compared with other cases. Fitting to a power law $\Delta E(L)=\Delta E(\infty)+\alpha L^{-\gamma}$, we obtain $\Delta E(\infty)=-4.1\,$meV, $\alpha=18.3\,$meV and $\gamma=0.68$. Note that the band gap at $\nu=0$ depends only weakly on $L$ since there is a direct gap at $\Gamma_{\text{M}}$. Hence in the NLSM picture, the $L$ dependence is expected to predominantly arise from the system size constraint on $R_\text{s}$ which controls the interaction contribution to Eq.~\ref{eq:NLSM}. Given that the gate screening length is comparable to the moir\'e length $d_\text{sc}\simeq a^\text{M}$, one expects $\gamma>1$ in the continuum limit for $R_\text{s}\gg a^\text{M}$. The discrepancy suggests that lattice corrections (the dotted patterns in Fig.~\ref{fig:1eskyrmions}a denote the $AA$-stacking regions) and finite size effects have a quantitative impact here. 

In the presence of anisotropies, the $1e$-skyrmions become finite-sized. For example in the chiral-nonflat limit with $J\neq0,\lambda=0$, the charge density is localized to within a few moir\'e lengths, leading to faster energy convergence with system size. This is consistent with the intuition from the NSLM, and the energetics of the skyrmion becomes significantly less favorable. The parent insulator is a member of the $SU(2)$ family containing the KIVC and the valley Hall state (e.g. polarizing into $KA$ and $K'A$ bands). The global net Chern polarization is less than $1/L^2$, consistent with the fact that the superexchange tunnels between opposite Chern sectors. 

A similar story holds for the nonchiral-flat limit with $J=0,\lambda\neq0$, except the $\nu=0$ insulator now interpolates between the KIVC and the valley-polarized state. This time, the added skyrmion is perfectly Chern polarized.

Including both perturbations, which is the case for realistic TBG, the symmetry reduces to the $U(1)$ family of KIVC insulators, and the Chern polarization of the skyrmion is imperfect again. Fig.~\ref{fig:1eskyrmions}d illustrates that the localized violation of Chern anti-alignment occurs at the same position as the charge density modulation, confirming the skyrmionic nature of the added charge. In Fig.~\ref{fig:1eskyrmions}c, we chart the relative energy of the skyrmions as a function of $\theta$ for different chiral ratios. For the average scheme, the $1e$-skyrmions are actually most costly near the magic angle where $J$ is supressed (compare with $\ref{fig:Jlambrhos}$b). This is despite the fact that artificially turning off $J$ while keeping other parameters fixed improves the skyrmion energy significantly (Fig.~\ref{fig:1eskyrmions}b). On the other hand, $\lambda$ is largely constant as a function of $\theta$. This suggests that the continuum description in terms of a small number of coupling parameters is not completely adequate. 

The numerical results for $\Delta E$ in any plot such as Fig.~\ref{fig:1eskyrmions}c are generally expected to represent upper bounds for two reasons. First, our calculations are performed on finite system sizes $L$. In the thermodynamic limit $L\rightarrow\infty$, the skyrmions will have some ideal radius $R_\text{s}(\infty)$ set by the intrinsic properties of the BM model and interaction potential. Unless $L\gg R_s(\infty)$, the pseudospins in our calculation will experience some degree of frustration from the finite simulation cell, leading to an energy penalty. In addition, a larger $L$ introduces a greater number of basis Bloch states which allows for smoother pseudospin rotations. Second, the restriction to Slater determinants in HF likely impacts skyrmions more than particle-like excitations. As shown in Sec.~\ref{subsec:effective_mass}, skyrmions can gain a small delocalization energy by going beyond mean-field and restoring the translation symmetry.

\subsection{$2e$-skyrmions}\label{subsec:2eskyrmions}

\begin{figure*}
	\includegraphics[ width=0.8\linewidth,clip=true]{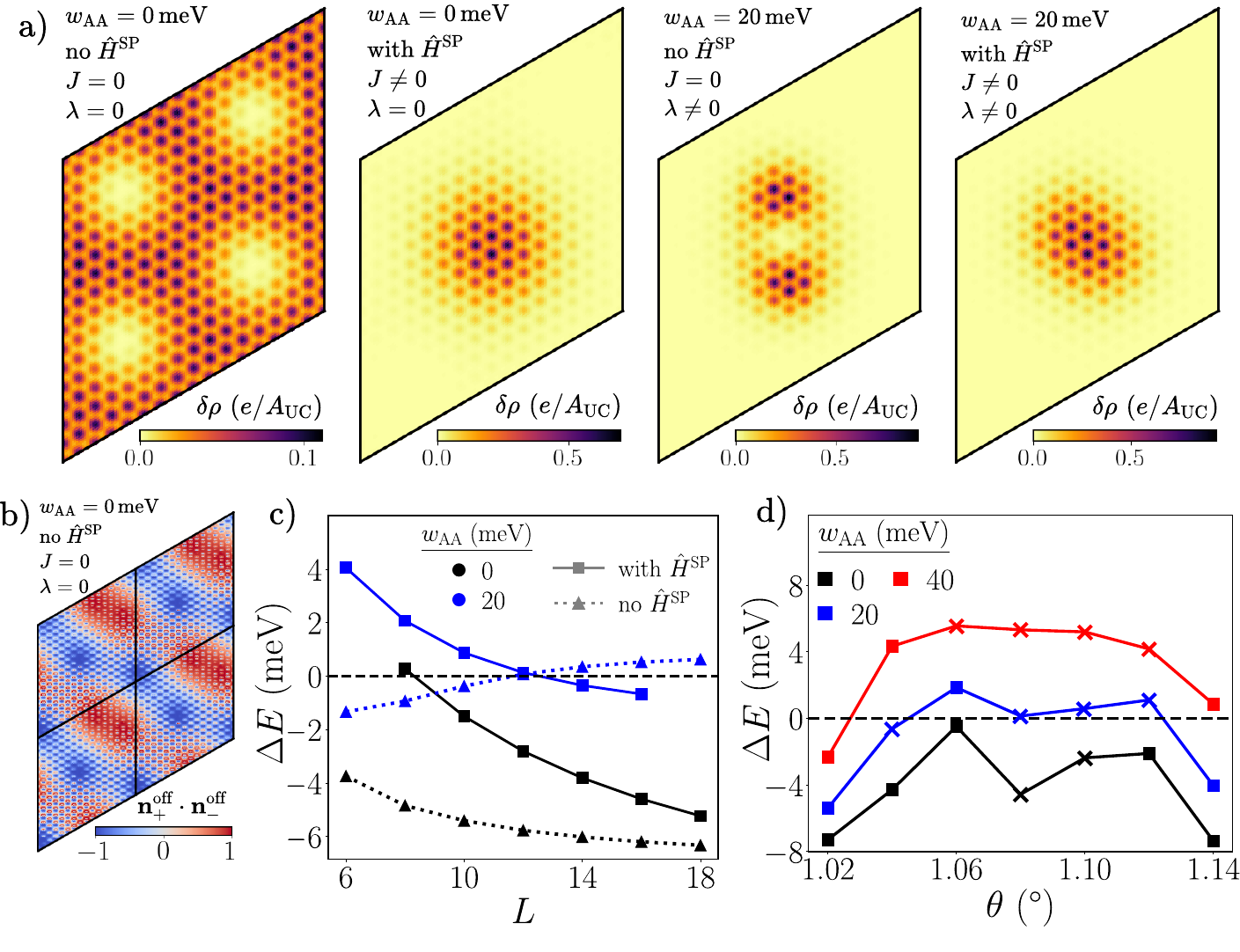}
	\caption{\textbf{Properties of $2e$-skyrmions (spinless $\nu=0$).} a) Charge density $\delta \rho(\bm{r})$ of $2e$-skyrmions in the entire simulation cell relative to $\nu=0$ ground state. Panels show different combinations of turning on/off dispersion and achirality, which determine $J$ and $\lambda$. Calculations performed on $16\times16$ systems at $\theta=1.05^\circ$. Textures manually translated to the center for clarity. b) Local alignment of Chern pseudospins corresponding to the leftmost panel of a). The system cell has been duplicated three times for presentation. c) Relative energy $\Delta E=E_{2e\text{-skyr}}-E_{2e}$ of $2e$-skyrmions (compared to adding two particle excitations) as a function of system size. $\Delta E<0$ indicates that the $2e$-skyrmion is favored. d) $\Delta E$ for different $\theta$ and $w_{AA}$. Dispersion is always included. Note that most of the solutions (indicated with crosses rather than squares) for $1.05^\circ<\theta<1.13^\circ$ are not $\hat{\mathcal{T}}'$-symmetric paired skyrmions. System size is $11\times11$. All calculations used the average scheme (see App.~\ref{app:BM_HF}).
	}
	\label{fig:2eskyrmions}
\end{figure*}

Fig.~\ref{fig:2eskyrmions}a shows the textured configurations when two electrons are added to the translation-invariant $\nu=0$ ground state. As before, the Chern couplings $J,\lambda$ can be controlled through the effective one-body term and the chiral ratio respectively. 

For $J=\lambda=0$, the texture breaks moir\'e translation symmetry completely, but the charge density appears to spread throughout the system without a clear identification of one or two well-defined objects. Interestingly, the lowest energy solution consists of both additional charges entering the same Chern sector, meaning that from a symmetry standpoint the situation will again be qualitatively similar to adding two charges to the QAH state. As shown in Fig.~\ref{fig:2eskyrmions}b, the pseudospin alignment (note that one Chern sector has a constant pseudospin) rotates in a complicated fashion without reconstructing a new superlattice periodicity. However in analogy to the case of $SU(2)$-invariant spin textures in Sec.~\ref{sec:spin}, the inclusion of more doped electrons can result in a so-called `double-tetarton lattice'~\cite{bomerich2020skyrmion} with emergent periodicity if they all Chern polarize. The resulting pseudospins form an ordered pattern as shown in Fig.~\ref{fig:spin_textures}a-c.

In the chiral non-flat limit with $J\neq0,\lambda=0$, the charge density consists of a single smooth modulation with approximate circular symmetry. This is precisely an explicit realization of a \emph{paired $2e$-skyrmion}: a skyrmion and an antiskyrmion with identical spatial profiles in the two Chern sectors exactly overlapping. This binding is induced by the superexchange $J$, as evidenced by the perfect anti-alignment of Chern pseudospins (and hence $\hat{\mathcal{T}}'$-symmetry). In the absence of other perturbations, the paired skyrmions spreads out and is limited only by the system size, leading to slow convergence in Fig.~\ref{fig:2eskyrmions}c. Hence the resulting physics controlling the texture is similar to that of the $1e$-skyrmion in the isotropic limit. 

In the nonchiral-flat limit with $J=0,\lambda\neq0$, one may na\"ively expect a similar paired skyrmion where the pseudospins are locked instead as $n_{+,z}=n_{-,z}$ and $\bm{n}_{+,xy}=-\bm{n}_{-,xy}$. However this does not work as it leads to an electrically neutral object. The numerics reveal that both charges predominantly go into the same Chern sector. For small system sizes, the state closely resembles the double-tetarton lattice of the isotropic limit, which is reflected in the energetic trends in Fig.~\ref{fig:2eskyrmions}c for small $L$. For larger sizes, the lattice deforms such that the texture is better described by a nearby pair of $1e$-skyrmions. While this may be considered pairing (actually since $\Delta E>0$, it is a metastable bound state that is unstable towards decaying into two particles), we reserve the term `paired skyrmion' for the $\hat{\mathcal{T}}'$-symmetric cases where a skyrmion and an antiskyrmion from opposite Chern sectors bind. It is noteworthy that $\Delta E$ increases as a function of $L$, bucking the trends of all other types of $1e$- and $2e$-skyrmions. This occurs because for $L\lesssim R_\text{s}(\infty)$, the constituent skyrmions are forced to strongly interact with each other in the confined system area
, and may therefore form a delocalized configuration with better energetics.

\begin{figure}
	\includegraphics[ width=1\linewidth,clip=true]{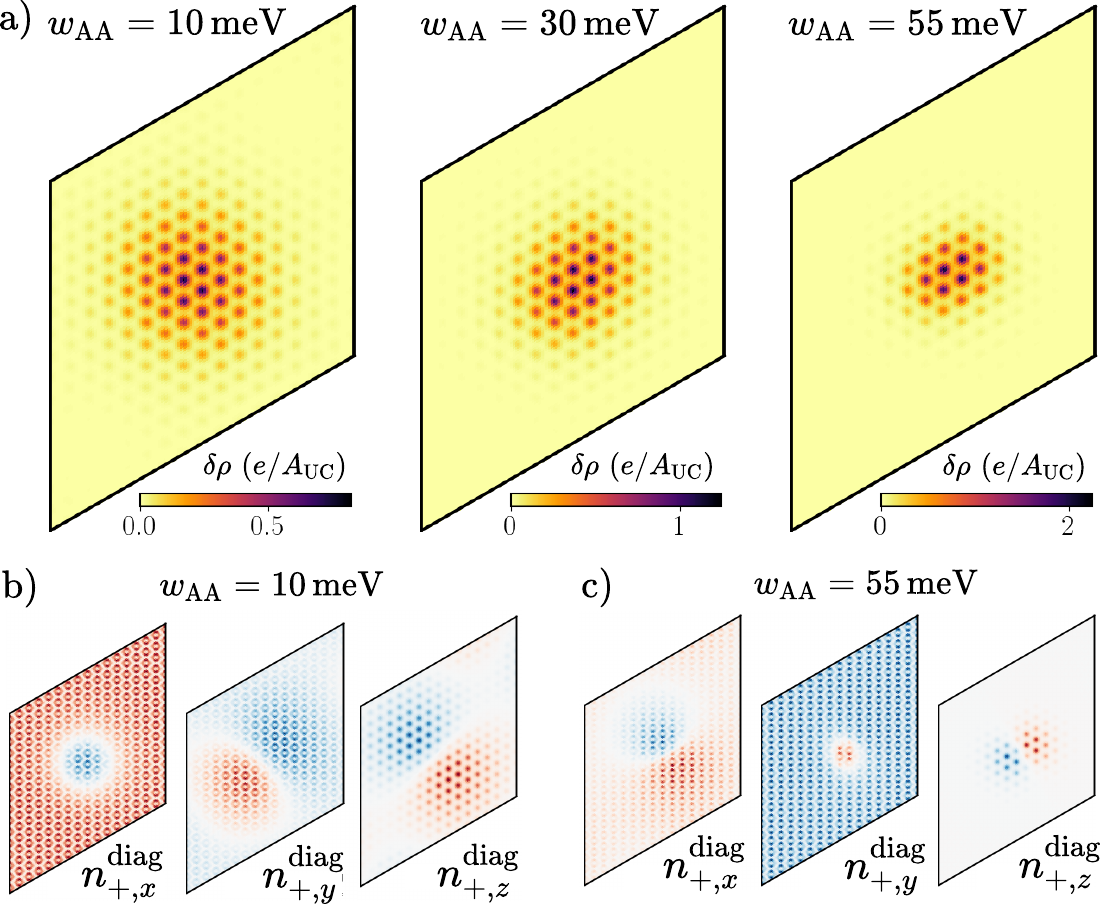}
	\caption{\textbf{Paired $2e$-skyrmions (spinless $\nu=0$).} a) Charge density $\delta\rho(\bm{r})$ of paried skyrmions for different $w_{AA}$. b,c) Local pseudospin orientation in one Chern sector corresponding to first and third panels of a). Note the increasing degree of anisotropy and confinement for increasing chiral ratio. System size is $16\times16$ with average scheme.}
	\label{fig:elliptical}
\end{figure}

It is clear from Fig.~\ref{fig:2eskyrmions}c that reintroducing dispersion to the achiral-flat limit is favourable to the textures, which are once again paired skyrmions. The faster convergence of the relative energy with $L$ is a hint that these paired skyrmions are now finite in size. This is clear from Fig.~\ref{fig:elliptical}a, which illustrates that increasing the chiral ratio not only reduces the skyrmion area, but also leads to an elliptical shape. As discussed in Sec.~\ref{subsec:NLSM}, this could be anticipated from the NLSM analysis which predicts that the $\lambda$-term penalizes the pseudospins when they anti-align out-of-plane. This anisotropy is apparent in Fig.~\ref{fig:elliptical}b,c, where $n_{+,z}(\bm{r})$ has a much tighter profile than $n_{+,xy}(\bm{r})$. At mean-field level, the 
orientation of the $n_z$ lobes is very soft, with distinct HF solutions differing by $\lesssim 10\,\mu\text{eV}$.

\begin{figure}
	\includegraphics[ width=1\linewidth,clip=true]{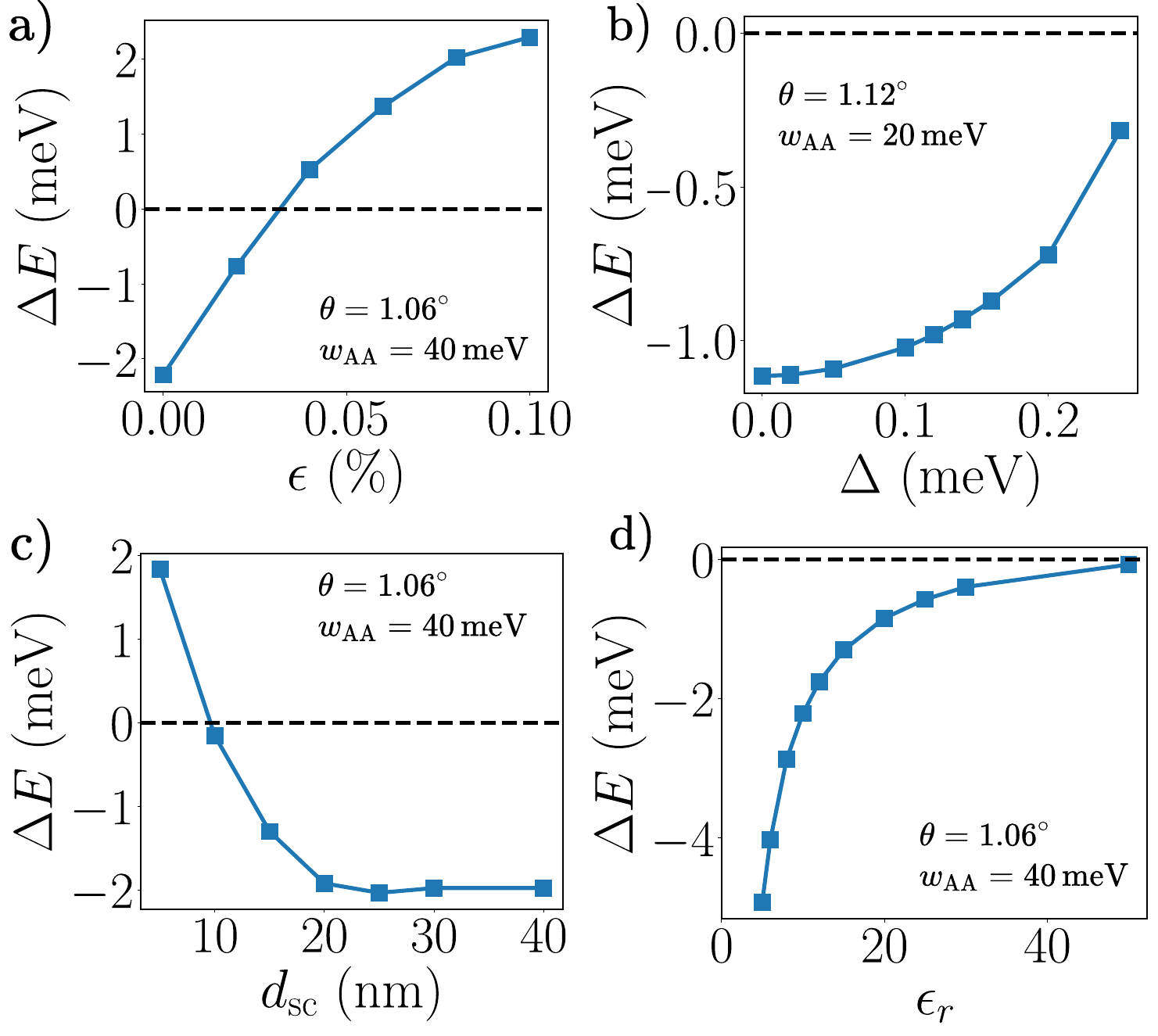}
	\caption{\textbf{Paired skyrmions and perturbations (spinless $\nu=0$).} Dependence of the relative energy of the paired skyrmion $\Delta E=E_{2e\text{-skyr}}-E_{2e}$ as a function of a) strain ratio $\epsilon$, b) substrate coupling $\Delta$, c) gate screening distance $d_\text{sc}$, and d) relative permittivity $\epsilon_r$. System size is $13\times13$ with graphene scheme. Qualitatively similar results are obtained for the other schemes.}
	\label{fig:perturbations}
\end{figure}

The energetic trends of the $2e$-skyrmions as a function of $\theta$ and $w_\text{AA}$ are shown in Fig.~\ref{fig:2eskyrmions}d. Note that specific to the average scheme, $\lambda$ significantly exceeds $J$ for a finite region of $\theta$ around the flat-band point. Consequently in the range $\sim1.05^\circ$ to $\sim1.13^\circ$, the best textured solution has finite Chern polarization and breaks $\hat{\mathcal{T}}'$, and hence is not a paired skyrmion.

Strain and substrate alignment represent two single-particle perturbations that are deleterious to the paired skyrmions, making them less energetically favorable and harder to find within HF. Strain significantly increases the kinetic bandwidth and is known to substantially degrade the gap of the KIVC at charge neutrality, eventually leading to a symmetric semimetal~\cite{parker2020straininduced}. As shown in Fig.~\ref{fig:perturbations}a, strain of strength $\epsilon\lesssim0.1\,\%$ is enough induce a positive relative energy $\Delta E$ for paired skyrmions. This should be interpreted in the context of STM studies which typically measure strains of $0.1-0.7~\%$~\cite{Kerelsky2019,Choi2019,Xie2019stm}. A staggered sublattice potential $\sim\Delta\sigma_z$ breaks $\hat{C}_{2z}$ and acts as a constant easy-axis Zeeman field with opposite direction in the two Chern sectors, which deters smooth rotations in pseudospin space (Fig.~\ref{fig:perturbations}b). For $\Delta$ beyond a fraction of an meV, HF fails to stabilize paired skyrmions at all. In fact if the coupling is strong enough, the mean-field ground state becomes a valley Hall state with no intervalley coherence~\cite{kwan2021kekule}. On the other hand, increasing the strength of interactions tends to favor the paired skyrmions. This can be achieved by increasing the gate screening distance $d_\text{sc}$ (Fig.~\ref{fig:perturbations}c) or reducing the relative permittivity $\epsilon_r$ (Fig.~\ref{fig:perturbations}d).

\subsection{Skyrmion composites and crystallization}\label{sec:composite}

\begin{figure*}
	\includegraphics[ width=1\linewidth,clip=true]{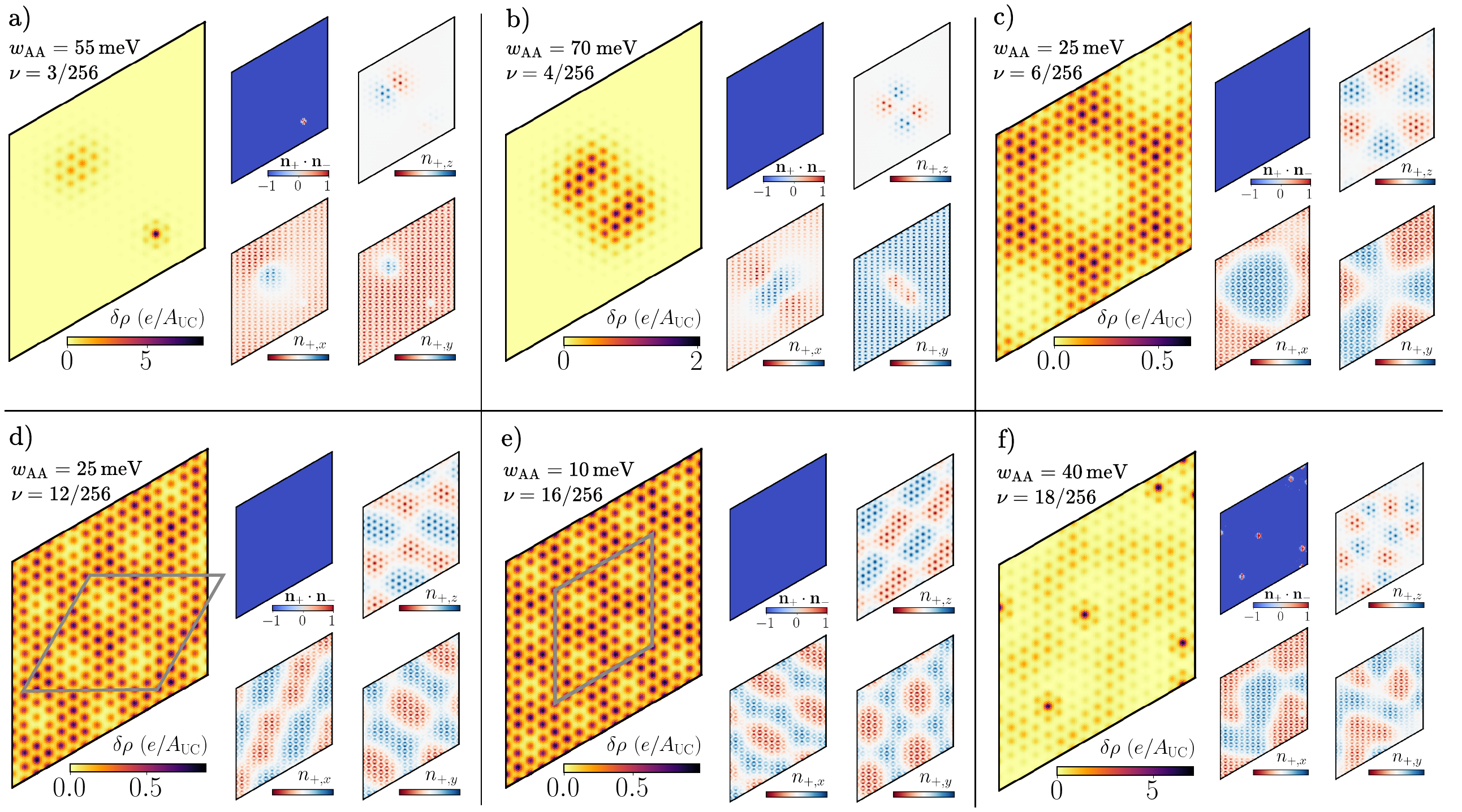}
	\caption{\textbf{Skyrmion composites and crystallization (spinless $\nu=0$).} Charge density $\delta\rho$, inter-Chern pseudospin alignment $\bm{n}_+\cdot\bm{n}_-$, and pseudospin orientation $\bm{n}_+$ for different numbers of electrons above charge neutrality $\nu=0$ in the spinless model. Emergent supercells for the double-tetarton lattice are highlighted in grey. All pseudospin densities are diagonal in microscopic layer/sublattice.  System size is $16\times16$ with average scheme.}
	\label{fig:composite}
\end{figure*}

By going beyond one or two particles, we can generate (metastable) higher skyrmion composites and skyrmion crystals. Fig.~\ref{fig:composite}a shows a typical configuration obtained by adding 3 electrons on top of $\nu=0$, resulting in a $1e$-skyrmion (in the $C=+1$ sector) and a paired $2e$-skyrmion that position themselves apart due to Coulomb repulsion. This identification can be straightforwardly made by looking at the pseudospin orientation. On the other hand, adding 4 electrons can lead to a $\hat{\mathcal{T}}'$-symmetric clump of two paired skyrmions which arrange their $n_z$ lobes in order to minimize the gradient cost (Fig.~\ref{fig:composite}b). For larger numbers of particles, the random initialization of the self-consistent HF loop tends to drive the system into local minima with complicated translation symmetry-breaking patterns (Fig.~\ref{fig:composite}f). Such states are often well-characterized by clumps or `trains' of paired skyrmions with suitably arranged pseudospin lobes, punctuated by $1e$-skyrmions within the gaps. The prevalence of such motifs, even for large $w_\text{AA}$ where particle-hole excitations are favorable, points to the robustness of skyrmions as well-defined localized excitations far away from the perturbative strong-coupling regime.

For intermediate dopings, the skyrmions can lose their individual identities and order into textured crystals. This was already shown in the isotropic limit for two added particles in Fig.~\ref{fig:2eskyrmions}a,b, but can occur even with anisotropies if the inter-skyrmion spacing becomes comparable with $R_\text{s}(\infty)$. For 6 extra particles, we can find a meron crystal where the lobes of the paired skyrmions lie on the honeycomb sites (Fig.~\ref{fig:composite}c, compare with Fig.~\ref{fig:spin_textures}d). Fig.~\ref{fig:composite}d,e are examples of double-tetarton lattices with different emergent supercells.  

\subsection{Effective mass of paired skyrmions}\label{subsec:effective_mass}
\begin{figure*}
	\includegraphics[ width=1\linewidth,clip=true]{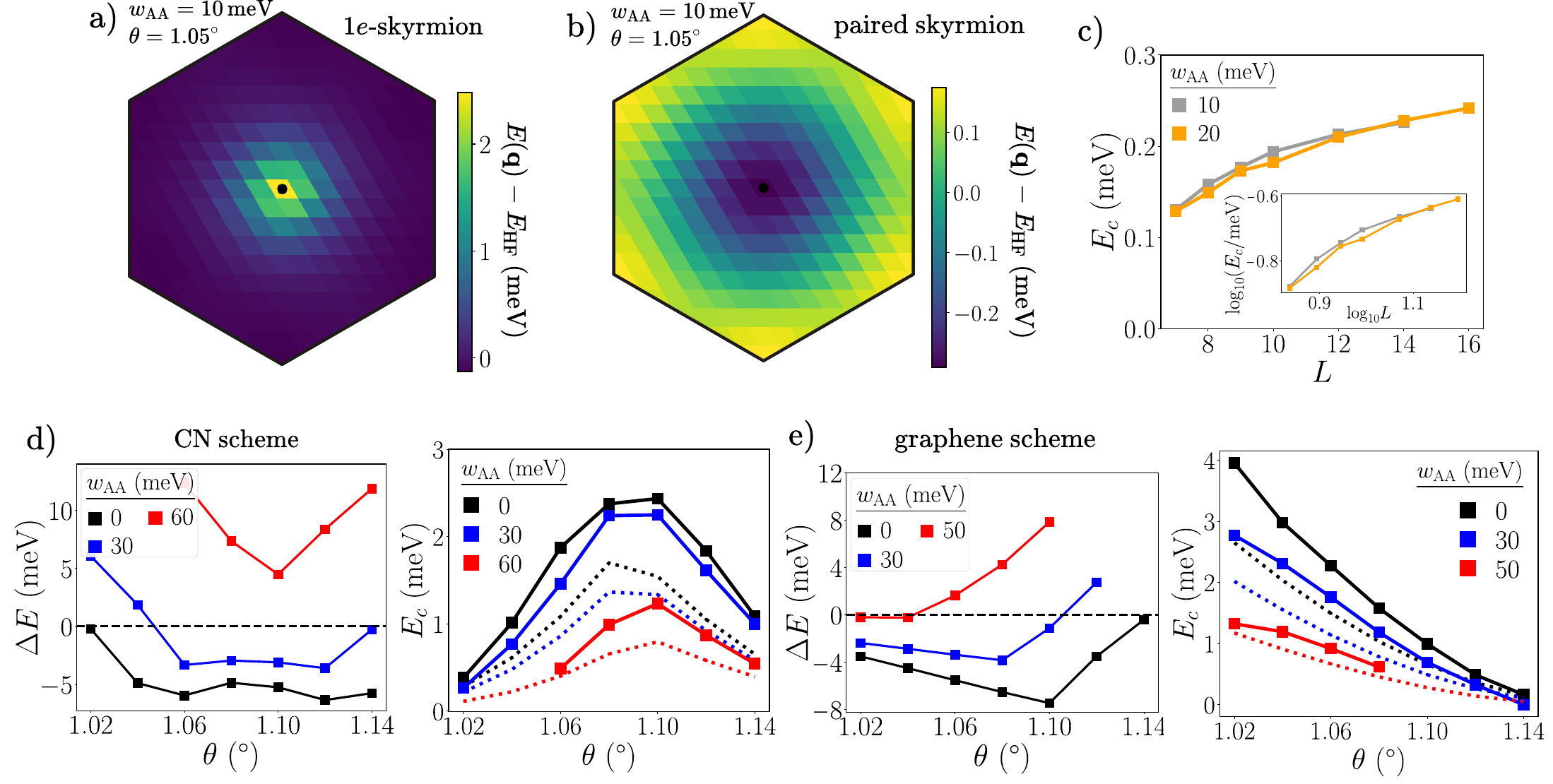}
	\caption{\textbf{Effective mass of paired skyrmions (spinless $\nu=0$).} a,b) Dispersion of $1e$-skyrmion and paired skyrmion respectively in the mBZ, obtained using an effective hopping model. Energy is measured relative to the localized HF solution. Black dot indicates $\Gamma_{\text{M}}$. System size is $14\times14$ with average scheme. c) Convergence of the superconducting BKT scale $E_c=\frac{\pi\hbar^2}{2A_\text{UC}M_\text{pair}}$ for the paired skyrmion with system size, in the average scheme. The effective mass $M_\text{pair}$ is fitted to the minimum of the skyrmion dispersion. Inset shows log-log plot. d) Left: Paired-skyrmion energy $\Delta E=E_{2e\text{-skyr}}-E_{2e}$ relative to particle excitation as function of $\theta$ for different $w_\text{AA}$ in the CN scheme. System size is $11\times 11$. Right: $E_c$ (solid) extracted from the skyrmion dispersion, compared with the superexchange scale $E_J/2$ (dotted) computed directly from the properties of $\hat{H}$. e) Same as d) except using a graphene scheme.
	}
	\label{fig:effective_mass}
\end{figure*}

A key prediction of the NLSM analysis is that paired skyrmions have a finite dispersion~\cite{khalaf2021charged}, which is crucial for generating a finite BKT energy scale $E_c$ for superconductivity. In the NLSM framework, this generation of a finite mass is non-trivial, arising from the interplay between the superexchange $J$ and the contrasting magnetic fields. An important question is how this picture holds up in the periodic moir\'e setting, where magnetic fields and flat bands are replaced by inhomogeneous Berry curavtures and interaction-renormalized dispersions. We address these issues using the effective hopping model described in App.~\ref{app:effective}, paying close attention to the dependence on the twist angle $\theta$ and subtraction scheme (the role of the subtraction is explained in Sec.~\ref{subsec:IBM} and App.~\ref{app:BM_HF}). 
Because of moir\'e translation invariance under $\hat{T}_{\bm{R}}$, our method is in essence a variational calculation of `Bloch skyrmions' $\ket{\psi_{\bm{q}}}=\frac{1}{L}\sum_{\bm{R}}e^{i\bm{q}\cdot\bm{R}}\hat{T}_{\bm{R}}\ket{\phi}$ based on a starting localized skyrmion $\ket{\phi}$ with an ideal pseudospin structure at the mean-field level. 
The required inputs are the matrix elements $\bra{\phi}\hat{T}_{\bm{R}}\ket{\phi},\bra{\phi}\hat{H}\hat{T}_{\bm{R}}\ket{\phi}$ of $\ket{\phi}$ and its translated images.

In Fig.~\ref{fig:effective_mass}a,b, we show the resulting bandstructure of $1e$-skyrmions and paired skyrmions, measured relative to the energy of the starting texture. The $1e$-skyrmion is characterized by a sharp peak at $\Gamma_\text{M}$ and shallow minima near the zone boundaries (the positions of these features may change for larger $w_{\text{AA}}$, but the overall structure remains the same). Usually a large number of overlaps and matrix elements beyond nearest neighbours needs to be computed to converge for a fixed set of parameters.   

On the other hand, the paired skyrmions are robustly associated with a broad energy minimum at $\Gamma_\text{M}$ and peaks at the zone corners. Typically only matrix elements for distances $\lesssim 2a^\text{M}$ need to be computed to accurately describe the properties of the full hopping model. The skyrmion mass $M_\text{pair}$, and hence the BKT energy scale $E_c$, can be estimated by fitting a parabola to the band minimum. This typically leads to $T_c$ of order $1$\,K. Fig.~\ref{fig:effective_mass}c plots the convergence of $E_c$ with system size $L$, which implies that the finite-size calculations here will typically underestimate the $L\rightarrow\infty$ results. As can be verified by checking nearest neighbor overlaps, there is no `orthogonality catastrophe' for moir\'e translations---the spatially inhomogeneous part of a paired skyrmion is finite in size, and most regions of the HF state remain translation invariant.

Having established the basic properties of the hopping model for paired skyrmions, we now turn to details of the dependence of $E_c$ on various parameters. The main motivation is to touch base with experiments which have observed a $T_c$ dome as a function of twist angle~\cite{Cao_2021}. However as noted in Sec.~\ref{subsec:2eskyrmions}, there is a large window of $\theta$ centred at the magic angle where paired skyrmions cannot be found in the average scheme. The utility of this scheme is that $J$ can be easily toggled by artificially turning off the kinetic Hamiltonian, but the drawback is that $J$ therefore becomes substantially smaller than $\lambda$ when the BM bands are flat, leading to a different type of $2e$-skyrmion. To address this issue, we consider now two alternative schemes which are not fine-tuned to have a vanishing superexchange.

\begin{figure}
	\includegraphics[ width=1\linewidth,clip=true]{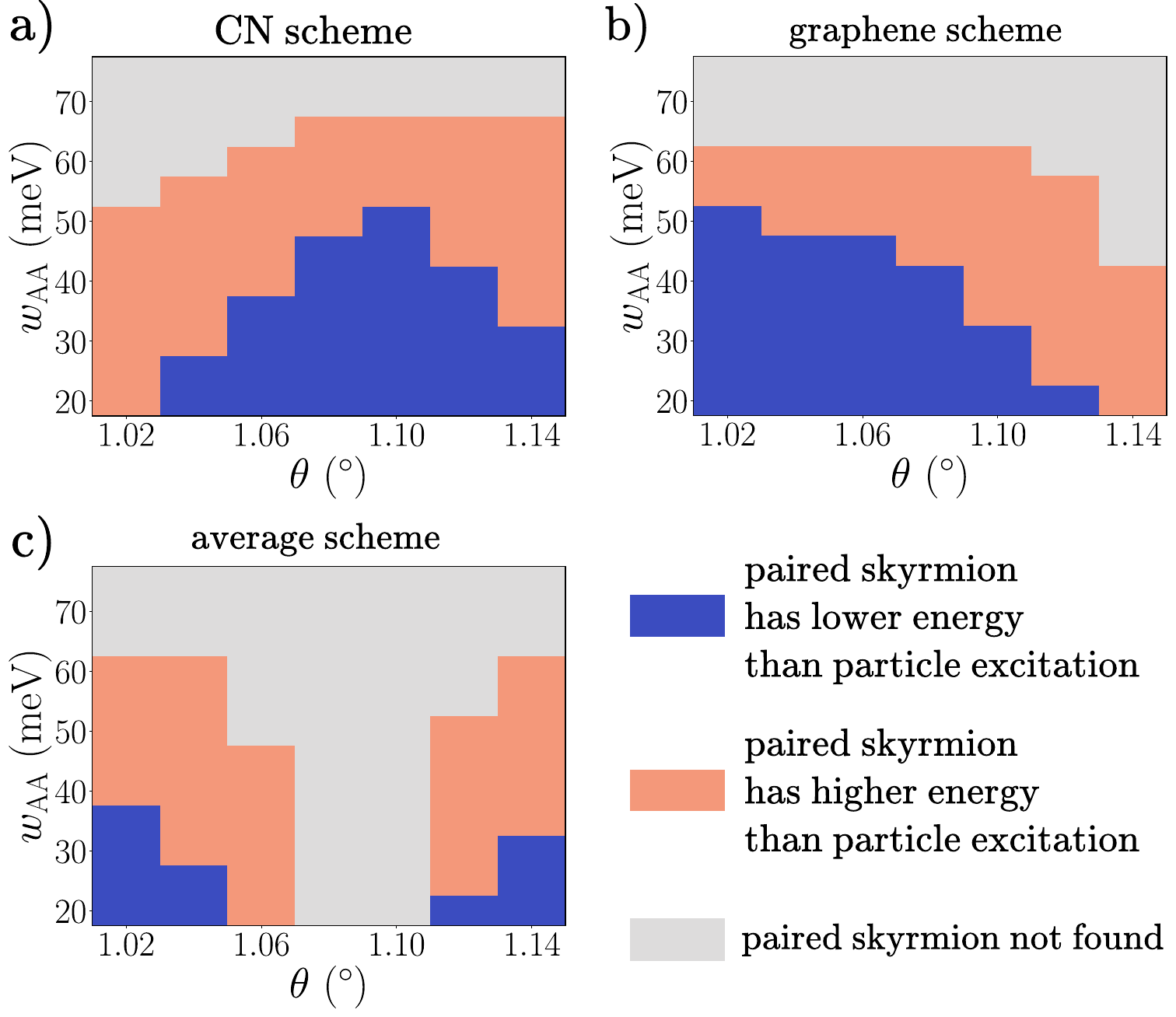}
	\caption{\textbf{Stability of paired skyrmions (spinless $\nu=0$).} a) Phase diagram in the $\theta$ vs $w_\text{AA}$ plane showing where paired skyrmions are energetically favorable (blue) or unfavorable (orange) compared to 2 particle-like excitations about the spinless $\nu=0$ insulator. Grey indicates regions where HF fails to find a metastable paired skyrmion solution. System size is $14\times 14$ with CN scheme. b,c) Same as a) but with graphene and average scheme.}
	\label{fig:phase}
\end{figure}

Fig.~\ref{fig:effective_mass}d charts the relative energy of the paired skyrmion as a function of $\theta$ and $w_\text{AA}$ for the CN scheme. In contrast to Fig.~\ref{fig:2eskyrmions}d, the paired skyrmions are energetically favored in an energy window centered around the magic angle. Again, $\Delta E$ has an inverted relation compared to $J$, which is plotted with dotted lines in the right panel. Note that paired skyrmions can be found up to $w_\text{AA}=60\,$meV for $L=11$, which should improve with increasing system size. Looking at the hopping model results, $E_c$ appears to qualitatively follow the same trends as $J$ for the whole range of twist angles investigated, suggesting that the physical intuition from the NLSM maintains some level of validity in the lattice case. Notably, the CN scheme suggests that a skyrmion superconductor would have a $T_c$ dome around the magic angle. 

The graphene scheme in Fig.~\ref{fig:effective_mass}e paints a somewhat different picture---$E_c$ monotonically increases in concert with $J$ for decreasing twist angle. For smaller angles the applicability of this calculation will be cut off by the fact that the parent insulator is no longer the ground state. Similarly the paired skyrmions, at least for small chiral ratios, become less favorable compared to particle excitations. Hence there are still possibilities for the hopping model results in the graphene scheme to be consistent with a $T_c$ dome.  

The stability of paired skyrmions in the CN, graphene and average schemes is summarized in Fig.~\ref{fig:phase}. Energetically favorable skyrmions can be found up to $w_\text{AA}\simeq 50\,\text{meV}$ for the right twist angles in the CN and graphene schemes (this is likely an underestimate when accounting for skyrmion delocalization and finite system size). Beyond this, metastable solutions are obtained for chiral ratios as large as $w_\text{AA}\simeq65\,\text{meV}$. Closer to realistic values of $w_{\text{AA}}\simeq 80\,\text{meV}$, we find that HF is not able to converge to any paired skyrmion states. In the average scheme, the paired skyrmions are energetically unfavourable close to the magic angle.

It is notable that different subtraction schemes lead to radically different behaviors in $E_c$. Indeed we believe that this is one of the few cases where such a choice impacts a physical quantity in a qualitative way. Many theoretical studies focus primarily on the type of symmetry-breaking order~\cite{Xie2020,XieSub,Bultinck_2020,2020CeaGuinea,Zhang2020HF,KangVafekPRL,Kang2020,Liu2021nematic,TBG4,TBG6,SoejimaDMRG,PotaszMacDonaldED,liu2021theories,shavit2021theory,zhang2021correlated,kwan2021kekule}---all the schemes studied here lead to the KIVC which only requires that $J,\lambda>0$. On the other hand, we are interested in the $\theta$-dependence of $E_c$ which depends sensitively on the value of $J$ itself.

The origin of this discrepancy can be understood by examining the effective dispersion $\hat{H}^\text{SP}$ in more detail. As explained in App.~\ref{app:BM_HF}, its matrix elements take the form
\begin{equation}\label{eq:hsp}
    h^\text{SP}(\bk)=h^\text{BM}(\bk)+\frac{1}{2A}\sum_{\bq}V_{\bq}\Lambda_{\bq}(\bk){Q^0}(\bk+\bq)^T\Lambda_{\bq}^\dagger(\bk).
\end{equation}
$J$ depends quadratically on the overall magnitude of $h^\text{SP}$. In the expression above, $P^0=\frac{1}{2}(1+Q^0)$ is the reference projector of the particular scheme. For the average scheme, $Q^0=0$ so that the only single-particle contributions (when the interaction terms are recast in strong-coupling form) arise from the kinetic piece $h^\text{BM}$ which gets heavily suppressed at the magic angle. 

$P^0$ for the CN scheme is constructed by occupying the valence bands of the BM Hamiltonian, while $P^0$ for the graphene scheme is built by filling the decoupled Dirac cones of each layer to the charge neutrality point and projecting to the central bands. An important clue is that both schemes qualitatively agree for $\theta$ above the magic angle. We focus on the states in the vicinity of the Dirac points for simplicity. For large angles, the Dirac cones of TBG are simply renormalized versions of those in decoupled graphene. In particular, we can calculate the behavior of the relative sublattice phase $\omega=\text{arg}(u_{\text{1A}}/u_{\text{1B}})$ for the valence band Bloch function as we traverse a small circle around a Dirac point. From a fixed starting point, this will wind by $2\pi$ with some initial offset that agrees for both TBG and decoupled layers. As $\theta$ is reduced, the BM bands distort until we reach the magic angle regime where the Dirac velocity vanishes and the bandwidth becomes tiny. Continuing past this point, the valence and conduction bands of TBG actually swap roles~\cite{Hejazi2019}, which is reflected in an additional $\pi$ offset in $\omega$.

The crucial insight is that the filling of the CN projector tracks this band reversal, while the graphene projector is oblivious to this physics. Note that the second term of Eq.~\ref{eq:hsp} represents the subtraction of the exchange gain for the filled bands of $P^0$. For the CN scheme, this is a positive contribution for the valence band of the BM model, which counteracts the negative contribution from the BM kinetic term itself. Therefore the two parts of Eq.~\ref{eq:hsp} tend to cancel each other out, leading to a dome at the magic angle when $h^\text{BM}(\bk)$ becomes suppressed. For the graphene scheme, this destructive interference occurs above the magic angle. However when the BM bands swap roles, the graphene projector does not follow, and hence the two parts add constructively for smaller $\theta$.

We reserve judgment on the matter of which scheme is most appropriate for capturing the physics in experimental TBG. Each choice has its own merits and justifications. The average scheme is the simplest and puts the strong coupling hierarchy front and center. The graphene scheme aims to prevent additional renormalizations of the Dirac cones that have already been accounted for in the input value for the bare Dirac velocity. The CN scheme provides a basepoint (i.e. charge neutrality of the BM model) at which the BM kinetic energy is precisely the mean-field band structure. However it is known to lead to incommensurate Kekul\'e spiral (IKS) order at extremely small strains for non-zero integer $\nu$~\cite{kwan2021kekule}. The microscopically correct answer is likely complicated and may differ based on the twist angle itself. 

Since paired skyrmions also break TRS and all point group symmetries, a natural extension of the effective mass calculation is to include symmetry-related partners of the starting HF solution $\ket{\phi}$ in a `hybridization $+$ hopping' model (see App.~\ref{app:effective} for details). By restoring the symmetries, this would shed light on the internal structure of the quantum skyrmion, including the angular momentum of pairing. However we find that an orthogonality catastrophe prevents feasibility of this method. Partners that are constructed from symmetries that do not leave the $\nu=0$ KIVC invariant, such as $\hat{\mathcal{T}}$ 
, have effectively vanishing overlaps with $\ket{\phi}$. This is not surprising since these operators have a non-trivial action on the pseudospins even in the bulk of $\ket{\phi}$ away from the localized paired skyrmion. However partners related by certain symmetries of the KIVC, such as $\hat{C}_3$, also have suppressed matrix elements and overlaps with $\ket{\phi}$. Fundamentally the reason is that the skyrmion texture is really a many-body pseudospin rotation of the parent insulator involving many particles, as evidenced by Fig.~\ref{fig:elliptical}b,c. Hence the Hamiltonian, which only involves few-body terms, cannot effectively couple the different partners.

\subsection{Spinful KIVC at $|\nu|=2$}

\begin{figure}
	\includegraphics[width=1\linewidth,clip=true]{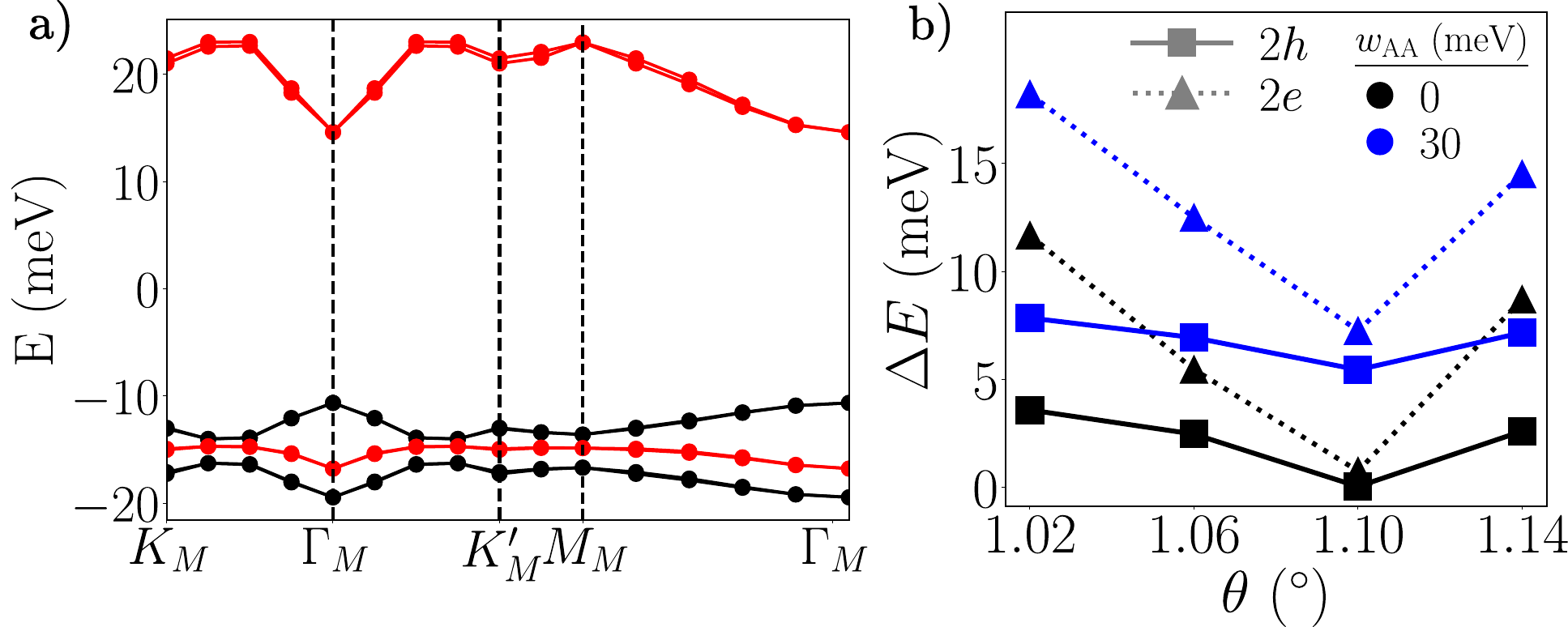}
	\caption{\textbf{Paired skyrmions (spinful $\nu=+2$).} a) Hartree-Fock dispersion of spin-polarized KIVC insulator at $\nu=+2$. Red (black) indicates minority (majority) spin carrier. System size is $24\times24$ with $\theta=1.06^\circ$. b) Relative energy $\Delta E=E_{2e(h)\text{-skyr}}-E_{2e(h)}$ of $\hat{\mathcal{T}}'$-symmetric paired $2e$- and $2h$- skyrmions about $\nu=+2$. Note sharp minimum of unfilled band at $\Gamma_\text{M}$. System size is $11\times 11$ with CN scheme.}
	\label{fig:sKIVC}
\end{figure}

We now reintroduce spin and turn to filling factors near $|\nu|=2$ where the mean-field ground state is also a KIVC insulator (we only show results for $\nu=+2$, but the situation at $\nu=-2$ can be inferred from PHS). A representative member of the spin-degenerate manifold of states, arising from the approximate $SU(2)_K\times SU(2)_{K'}$ symmetry, consists of fully filling the spin-$\uparrow$ flat bands while forming a spinless KIVC in the (minority) spin-$\downarrow$ subspace. 
We concentrate on pseudospin skyrmions by enforcing collinear spins. 

Paired skyrmions can be constructed as before by treating the majority spin bands as spectator bands and performing non-trivial pseudospin rotations in the half-filled minority spin bands, but there are complications compared with the spinless neutrality case. First, the NLSM parameters $J,\lambda$ reflect the relevant energy scales at the neutrality point of the strong-coupling Hamiltonian. However the starting insulator now contains additional majority carriers, which impacts the effective energetics of the minority subspace. Second, this additional interaction renormalization enters in a particle-hole asymmetric way. Generally the bands away from neutrality have a significantly enhanced bandwidth with a prominent extremum at $\Gamma_\text{M}$~\cite{kang2021cascades,pierce2021unconventional} (Fig.~\ref{fig:sKIVC}a), which disfavors skyrmions because they are built from momentum states throughout the mBZ. Third, the edge of the majority spin bands can lie inside the minority band gap which reduces the gap to particle-hole excitations. On the other hand, a pseudospin skyrmion cannot be formed by adding holes to a completely filled spin sector.

These considerations are reflected in the HF calculations in Fig.~\ref{fig:sKIVC}b at $\nu=+2$, showing that paired $2e$-skyrmions on the electron-doped side are more expensive than paired $2h$-skyrmions on the hole-doped side. Overall the relative energies $\Delta E$ are less favorable than the results at charge neutrality.

\section{Spin textures at $|\nu|=3$}\label{sec:spin}

\begin{figure}
	\includegraphics[ width=1\linewidth,clip=true]{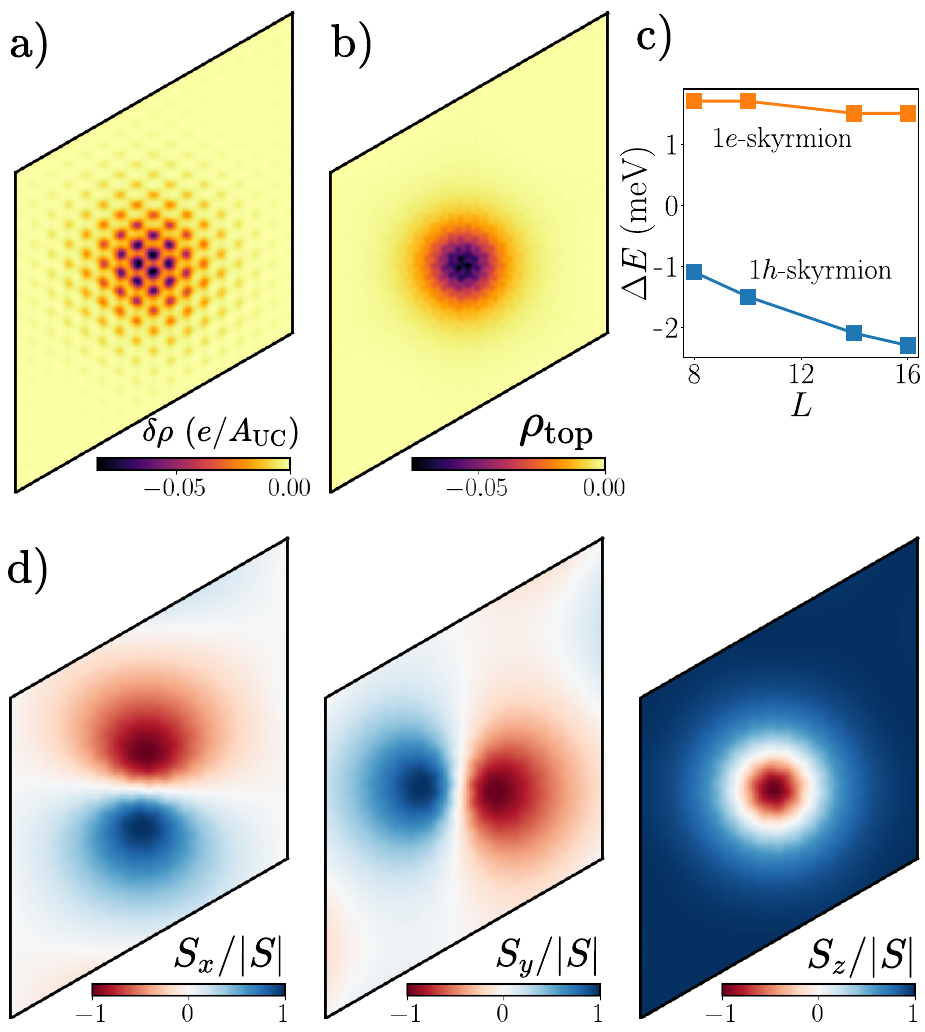}
	\caption{\textbf{Spin skyrmions about QAH (spinful $\nu=+3$).} a) Charge density of a single hole about the $\nu=+3$ QAH state. b) Topological density $\rho_\text{top}$ computed from the spatial spin configuration of a). c) Energy of single electron and hole skyrmions $\Delta E=E_{1e(h)\text{-skyr}}-E_{1e(h)}$ (relative to a particle-like electron or hole excitation) as a function of system size $L$. d) Local spin orientation $\bm{S}(\bm{r})/|\bm{S}(\bm{r})|$ corresponding to the skyrmion in a). The spins have been rotated so that the state has a net polarization along $z$. System size is $16\times16$ with graphene scheme.}
	\label{fig:spin_skyrmion}
\end{figure}

The physical intuition and considerations behind skyrmions in Sec.~\ref{sec:even_integer} can also be applied to odd integer fillings. In this section, we briefly discuss textures near $\nu=+3$ that arise from doping the Chern, spin and valley-polarized QAHI (the results at $\nu=-3$ can be found from particle-hole symmetry about neutrality). The situation here is significantly simpler because of the $SU(2)_S$ spin-rotation symmetry that holds independent of the presence of dispersion or deviation from the chiral limit. Furthermore the starting insulator is easy-axis in valley space. The low-energy charged topological excitations are then expected to be spin skyrmions~\cite{chatterjee2020symmetry,zhang2019nearly,bomerich2020skyrmion} in the partially filled Chern sector without any texturing in the fully filled Chern sector.

In Fig.~\ref{fig:spin_skyrmion}, we compute the properties of $1h$- and $1e$-skyrmions by relaxing the spin collinearity constraint in our HF calculations. In a similar fashion to isotropic $1e$-skyrmions in Fig.~\ref{fig:1eskyrmions}a, the change in charge density is roughly circularly symmetric with moir\'e-periodic modulations on the $\text{AA}$-stacking regions. The local spin orientation is consistent with the hallmark features of an $SU(2)$-symmetric skyrmion. Indeed the calculation of the topological density $\rho_\text{top}$ (in spin space) confirms the topological origin of the excitation. Unlike in QHFM, there is no explicit Zeeman field, meaning that the skyrmion expands to fill the available system size to minimize the Coulomb energy. This is reflected in the slow convergence of $\Delta E$ with $L$. There is a clear asymmetry between adding holes and electrons because the interaction-renormalized band structure~\cite{Guinea2018,rademaker2019smoothening,cea2019pinning,goodwin2020hartree,kang2021cascades,pierce2021unconventional} at finite integer fillings strongly breaks particle-hole symmetry (see Fig.~\ref{fig:sKIVC}a).

\begin{figure}
	\includegraphics[ width=1\linewidth,clip=true]{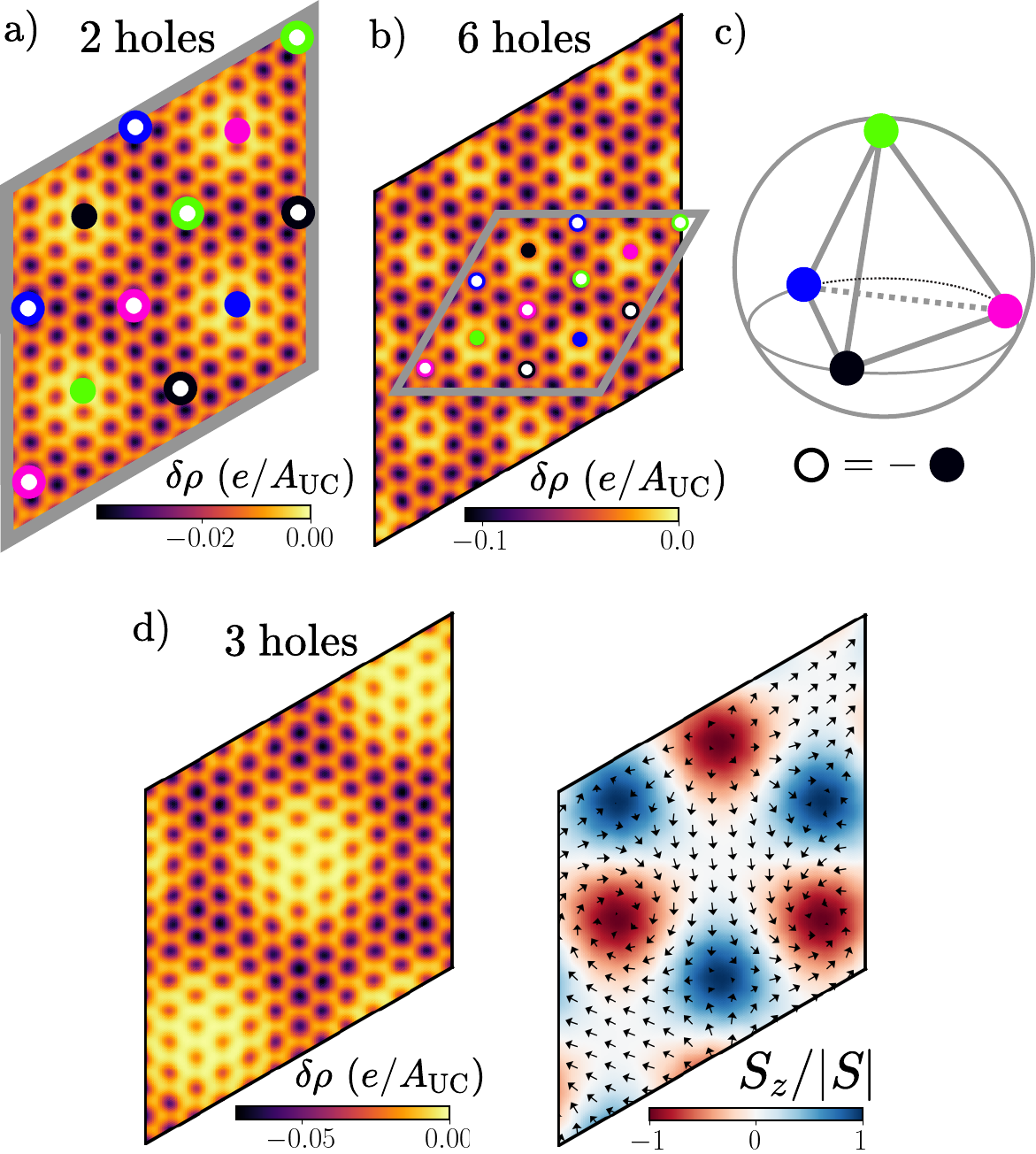}
	\caption{\textbf{Spin skyrmion crystallization about QAH (spinful $\nu=+3$).} a,b) Charge density when adding 2 and 6 holes respectively. Grey parallelogram indicates the supercell. c) In the double-tetarton lattices of a,b), the spins at particular positions (anti)align along the corners of a tetrahedron in spin space. d) Charge and $z$-component of local spin orientation when adding 3 holes, leading to a meron lattice. System size is $12\times12$ with graphene scheme.}
	\label{fig:spin_textures}
\end{figure}

Just as in Sec.~\ref{sec:composite}, doping additional charges leads to the formation of various skyrmion crystals. The unit cell of the double-tetarton lattice~\cite{bomerich2020skyrmion} contains two charges with an intricate spin pattern illustrated in Fig.~\ref{fig:spin_textures}a,c. An alternative configuration is the meron crystal whose unit cell is associated with a charge of $3e$ (Fig.~\ref{fig:spin_textures}d). Computations on system sizes and dopings where both options can be stabilized are required to numerically determine the energetically preferred lattice. It has also been proposed that applying a Zeeman field will enrich the possible set of phases~\cite{bomerich2020skyrmion}.

\section{Discussion}\label{sec:discussion}
In closing, we consider various generalizations and extensions of the work presented above, comment on the possibility of experimentally observing skyrmion physics, and critically assess the implications of our work for the proposed skyrmion-driven mechanism for superconductivity in TBG.
\subsection{More general skyrmions}
In this work, motivated by the particular physics and questions relevant to TBG, we have imposed restrictions on the directions in flavor space that the skyrmions are allowed to rotate in. The strong-coupling insulators at different integer $\nu$ are all Chern ferromagnets, and starting from the fully symmetric limit, one can consider more general skyrmions where the only constraint is that locally the state is polarized within the Chern sectors. For instance, starting from the $|\nu|=+3$ QAHI, it is possible to form a texture in both spin and pseudospin to create an `entangled' skyrmion, akin to what happens at $|\nu|=1$ in the zeroth LL of monolayer graphene~\cite{lian2017spin}. With the full Hamiltonian, pseudospin rotations will be gapped since the QAHI is easy-axis, while spin rotations remain low-energy due to the $SU(2)_S$-symmetry.

These considerations can be generalized to any integer filling. Starting from a generalized ferromagnetic insulator with $\tilde{\nu}_C$ filled bands in Chern sector $C$, the local flavor configuration is parameterized by two matrix spinors $f_C(\bm{r})$ living in Grassmannian projective spaces~\cite{atteia2021skyrmion}
\begin{equation}
    f_C(\bm{r})\in \frac{SU(4)}{SU(\tilde{\nu}_C)\times SU(4-\tilde{\nu}_C)\times U(1)}
\end{equation}
where $\tilde{\nu}_++\tilde{\nu}_-=\nu$. The factors in the normal subgroup represent unitary rotations in within the filled bands, unfilled bands, and the phase difference between the filled and unfilled bands respectively. The presence of four flavors and two spinors leads to a large manifold of skyrmions. From the discussion in  Sec.~\ref{subsec:strong_coupling}, the leading corrections from the $U(4)\times U(4)$ limit will be anistropic \emph{couplings} between the spinors, while additional single-particle perturbations such as substrate coupling may manifest as anisotropic \emph{fields}. The relevant space of textures will be dictated by the energetics at a given filling. 

The ground state in the spinful model at charge neutrality is the spin-unpolarized KIVC~\cite{Bultinck_2020}
\begin{equation}
    Q_{\text{KIVC}}=\sigma_y(\tau_+V+\tau_-V^\dagger)
\end{equation}
where $V=e^{i\phi}e^{i\frac{\varphi}{2} \bm{m}\cdot\bm{s}}$ and $\tau_\pm=\frac{1}{2}(\tau_x\pm i\tau_y)$. This represents two copies of the spinless version in Eq.~\ref{eq:QKIVCspinless} with spin projections $\pm \bm{m}$ and IVC phases $\phi\pm\frac{\varphi}{2}$ (the relative phase $\varphi$ is set by the intervalley Hund's terms). Equivalently we have four Chern pseudospins $\bm{n}_{C,s}$ where $s$ indicates the projection of the spin along $\bm{m}$. Consider doping two electrons. Forming a paired skyrmion in one of the spin sectors in the usual way can be understood using the analysis of Sec.~\ref{sec:even_integer}. Attempting to rotate in spin space will either lead to Pauli blocking or the loss of antiferromagnetic exchange $J$, which justifies our restriction to pseudospin textures in Sec.~\ref{sec:even_integer}. Similar conclusions can be reached for the spin-polarized KIVC insulator at $|\nu|=2$, though we cannot rule out the majority spin being non-trivially involved if it is the closest filled band to the chemical potential.

\subsection{Disorder}
The effects of disorder on skyrmions have been treated previously in the quantum Hall context~\cite{nederveen1999,sinova2000,rapsch2002,lee2003,fertig2005}. A strong and smooth disorder potential can drive the system into a spin glass. We note that all the strong-coupling insulators of interest in this work violate TRS, and hence ordinary impurities cannot directly couple to the symmetry-breaking order as a random field. 
Isolated charged impurities are expected to pin individual skyrmions as well as spin texture lattices, which may aid in their detection as discussed in the following subsection. TBG can also harbor more subtle forms of disorder. For the QAHI, modulations in the magnitude and sign of the sublattice coupling $\Delta$ can lead to the enforced nucleation of domain walls, including ones which separate regions with opposite Chern number~\cite{Kwan2020domain}. These would interrupt the propagation and crystallization of spin skyrmions. Twist angle inhomogeneity~\cite{wilson2020}, which is pervasive in experimental samples~\cite{uri2020mapping,Kerelsky2019,Choi2019,jiang2019charge,Xie2019stm}, is notoriously difficult to model, and its theoretical impact on the correlated insulators at integer fillings is still beyond quantitative characterization. For sufficiently long-wavelength fluctuations, the properties of the skyrmions (e.g.~the effective masses) are likely to depend on the local twist angle.

\subsection{Detecting skyrmions}
The presence of spin textures about the QAHI at $\nu=+3$ could be detected by measuring the degradation of magnetization for small dopings, e.g.~through NMR measurements~\cite{barrett1995}, NV center magnetometry, or using a superconducting quantum interference device (SQUID)~\cite{tschirhart2021imaging}. Individual spin skyrmions involve a large number of spin flips, and the spin texture lattices of Fig.~\ref{fig:spin_textures} are close to completely spin unpolarized. However there is an orbital contribution from the spontaneous Chern polarization which is likely larger~\cite{tschirhart2021imaging}. A more direct probe would be spin-resolved STM near impurity sites that could pin a localized skyrmion.

Measuring pseudospin textures at even integer filling is trickier since experimental techniques are not able to directly couple to the valley degree of freedom. IVC generally leads to $\sqrt{3}\times\sqrt{3}$ spatial order at the microscopic graphene scale, but the $\hat{\mathcal{T}}'$-symmetry of the KIVC insulator means that it does not exhibit a Kekul\'e \emph{density} distortion (KD)~\cite{caluguru2021spectroscopy,hong2021detecting}. Paired skyrmions preserve $\hat{\mathcal{T}}'$ and hence do not give rise to KD~\cite{hong2021detecting}. They may still leave a dipole-shaped fingerprint in sublattice polarization within regions where the state is locally in the valley Hall configuration, i.e.~pseudospins anti-aligned and pointing along the $z$-axis (Fig.~\ref{fig:elliptical}c), but this is likely a faint signature since away from the chiral limit the Chern basis is only partially sublattice polarized. On the other hand, $1e$-skyrmions are tightly localized within a few moir\'e lengths and give rise to a spatially varying KD when pinned by charged impurities. KD has been observed in the related context of QHFM within the lowest LL of monolayer graphene~\cite{Li2019Kekule,liu2021visualizing,coissard2021imaging}, including the imaging of an individual valley skyrmion~\cite{liu2021visualizing}. We caution that KD in TBG has also been predicted for IKS order in the presence of a small amount of strain~\cite{kwan2021kekule,hong2021detecting}.

\subsection{Skyrmion superconductivity}

We finally turn to the implications of our work for skyrmion superconductivity.
 First, note that two important factors controlling the feasibility of this proposed mechanism are the stability of paired skyrmions (i.e.~relative energy $\Delta E$ compared to particle excitations) and their effective masses. In general, we can conclude that paired skyrmions are especially favored close to the strongly-interacting isotropic limit. The realistic value of $w_\text{AA}$ lies in the range $55-90\,\text{meV}$ (i.e.~$\kappa\simeq0.5-0.8$) and we only find paired skyrmions in the lower range of these values (Fig.~\ref{fig:phase}). We note that a mechanism has been proposed that might drive a downward renormalization of the chiral ratio $\kappa$~\cite{vafek2020RG}.
 We note also that large skyrmions, which are relatively classical and relevant for small $w_{AA}$, are likely to be well captured in our mean-field treatment. However, for larger $w_{AA}$, the paired skyrmions become smaller and quantization effects are more important, and fluctuations can be more significant. In this regime the mean-field result is really only an upper bound on the skyrmion energy, which could be lowered by fluctuations, which are not expected to substantially affect the single-particle excitations. 
 
 Enhancing interactions by suppressing  screening, either through increasing the gate distance $d_\text{sc}$ or decreasing the permittivity $\epsilon_{r}$, also favors skyrmions (Fig.~\ref{fig:perturbations}). However, this observation makes the superconducting domes that persist in Refs.~\onlinecite{saito2020independent,Stepanov_2020,liu2021tuning,oh2021evidence} upon reducing the interaction strength difficult to reconcile with a topological mechanism. In these experiments, the insulators at integer $\nu$, from which the skyrmions would be seeded, disappear with increasing screening. The superconducting region can also straddle the integers where the BKT transition temperature from Eq.~\ref{SF_Tc}  seemingly vanishes. Substrate coupling rapidly destroys  pseudospin skyrmions, consistent with the absence of superconductivity in aligned samples~\cite{Sharpe_2019,Serlin900}. This topological mechanism would not be effective in other moir\'e platforms that lack $\hat{C}_{2z}$-symmetry~\cite{khalaf2021charged,cao2020tunableTDBG,shen2020correlated,liu2020tunableTDBG,chen2019signatures}. Strain takes the system away from the strong-coupling regime and  similarly disfavors skyrmions: for instance, the parent insulator has been predicted to give way to a symmetric semimetal at charge neutrality or an IKS at non-zero integer fillings~\cite{parker2020straininduced,kwan2021kekule}. Hence the general expectation from our work and from these experimental observations is is that skyrmion superconductivity is most likely to emerge in `pristine' samples with minimal screening.

The question of effective masses (and hence $T_c$ via Eq.~\ref{SF_Tc}) is more subtle. We have demonstrated that the dependence of the BKT transition scale $E_c$ on $\theta$ is rather sensitive to precise details of how electron interactions are incorporated. Without further external inputs, e.g.~from detailed \textit{ab initio} studies or spectroscopic probes of the band structure over a range of twist angles, it is difficult to make quantitative contact with experiments such as Ref.~\onlinecite{Cao_2021} which show a dome in $T_c$ near the magic angle. Any comparison would also inevitably be complicated by the presence of confounding variables such as twist angle disorder~\cite{uri2020mapping} which are difficult to fully characterize, let alone control. However, what we \emph{can} reliably distill from our numerical study is that paired skyrmions can 
in principle emerge with a sufficient mass
to support an estimate of $T_c\lesssim 5\,\text{K}$  that is comparable to experimentally observed values. Both the CN and graphene schemes are able to support a non-vanishing superfluid velocity at the magic angle, and the fragility of the parent correlated insulators to deviations of $\theta$ will tend to also reduce the strength of skyrmion superconductivity away from this regime. 

The discussion above focuses primarily on a ``BEC--like'' limit of skyrmion superconductivity, where the binding is present already in the dilute limit; this is the regime primarily accessed in this work. An intriguing alternative possibility is a more ``BCS-like'' picture, where skyrmions while unstable at low density nevertheless become favored and paired at finite density. We note that Ref.~\onlinecite{chatterjee2020skyrmionSC} which studies skyrmion superconductivity in the Landau level limit found superconductivity in the doped case, even in the absence of a pairing gap at zero doping.

Another potential challenge to the applicability of the skyrmion mechanism to TBG lies in the fact that superconductivity is most frequently observed near $\nu=-2$ on the side \emph{away} from charge neutrality. On the other hand, our numerics show that paired skyrmions are relatively harder to stabilize at $|\nu|=2$ compared to $\nu=0$ (Fig.~\ref{fig:sKIVC}b). Furthermore, the skyrmions are more disfavored when doping in the direction away from charge neutrality. This latter observation can be explained by the increased dispersion due to the interaction renormalization from the extra filled bands (Fig.~\ref{fig:sKIVC}a), and has been argued to be consistent with the asymmetry of the Landau fans~\cite{kang2021cascades}.

We cannot rule out a scenario where  skyrmion superconductivity is operative in only a subset of samples, for instance the device studied in Ref.~\onlinecite{Lu2019} which is nominally non-aligned with hBN and exhibits an remarkably large number of correlated insulators and superconducting domes, including near neutrality. Another moir\'e material where the skyrmion mechanism may be a plausible explanation of superconductivity   is mirror-symmetric magic-angle twisted trilayer graphene (TTG), which is closely related to TBG but has a somewhat larger value for the magic-angle \cite{KhalafTTG}. Interestingly, the superconductor in TTG is observed to have a very short coherence length \cite{MATTGPablo}, and an associated pseudogap regime \cite{MATTGPerge}. Partly because of this, 
Refs.~\onlinecite{MATTGPablo,MATTGPerge} suggested that part of the TTG superconducting dome is in the BEC regime. Skyrmion pairing is a natural way to get preformed charge-2e bosons, and is at least known to give rise to superconductivity in the chiral limit of TBG \cite{chatterjee2020skyrmionSC}. It is therefore worthwhile to investigate whether this mechanism can  explain at least a subset of the experimental observations in TTG.

In summary, our work provides clear evidence that the formation and pairing of skyrmions can indeed occur in microscopically faithful treatments of TBG, thereby illustrating that a purely electronic ``topological''  Cooper pairing mechanism can operate 
  away from the exactly-solvable limit without leveraging any approximate sigma-model description. 
 However, despite this in-principle demonstration of the feasibility of a novel pairing mechanism, we cannot on the basis of present evidence definitively attribute superconductivity in TBG to this mechanism. This is  highlighted by the difficulty in  reconciling the deleterious effect of variations in strain and interaction strength and deviation from the chiral limit  on the stability of skyrmions with the apparent robustness of superconductivity to such effects. Despite this, the uncertainty of various microscopic parameters and the mean-field nature of our study leaves open a real possibility that a skyrmionic mechanism may ultimately survive these challenges. Further work, especially involving numerical approaches that can capture fluctuations beyond the mean-field level, is clearly warranted, as are the exploration of new regimes and systems where skyrmion pairing may be more favored and the identification of new probes that can directly interrogate the nature of the pairing `glue' in TBG.

\vspace{0.5cm}
\noindent\textbf{Note Added.} A preprint by Eslam Khalaf and Ashvin Vishwanath appearing in the same arXiv posting performs a complementary analysis of $1e$-skyrmion formation in the limit of small skyrmion size. In this regime, skyrmions may be viewed as ``spin polarons'', consisting of a charged electron or hole bound to a single spin/pseudospin flip particle-hole pair excitation. Our results are broadly in agreement where they overlap.

\begin{acknowledgments} 
 We thank Andrei Bernevig, Shubhayu Chatterjee, Eslam Khalaf, Ashvin Vishwanath, and Michael Zaletel for discussions. We are grateful to Eslam Khalaf and Ashvin Vishwanath for sharing an advance copy of their preprint (see Note Added) and to both of them and Mike Zaletel for comments on this manuscript. GW acknowledges NCCR MARVEL funding from the Swiss National Science Foundation. NB is supported by a senior postdoctoral research fellowship of the Flanders Research Foundation (FWO). We acknowledge support from the European Research Council under the European Union Horizon 2020 Research and Innovation Programme, Grant Agreement No. 804213-TMCS and  from EPSRC Grant EP/S020527/1. Statement of compliance with EPSRC policy framework
on research data: This publication is theoretical work
that does not require supporting research data.
\end{acknowledgments}

\bibliography{bib}
 
\clearpage
\newpage

\begin{widetext}
 \begin{appendix}

\section{Interacting BM model and Hartree-Fock}\label{app:BM_HF}
\subsection{BM Model}
Here we review the definition of the Bistritzer-MacDonald (BM) model for the case of unstrained TBG. Modifications relevant to the strained case are described in the Appendix of Ref.~\onlinecite{kwan2021kekule}.

The basis direct lattice vectors (LVs) and reciprocal lattice vectors (RLVs) for the unstrained TBG moir\'e pattern (where layers 1 and 2 are rotated counterclockwise by $\theta/2$ and $-\theta/2$ respectively) are chosen to be
\begin{gather}
    \bm{a}^\text{M}_1 =\frac{4\pi}{3k_\theta}(\frac{\sqrt{3}}{2},\frac{1}{2}),\quad\bm{a}^\text{M}_2 =\frac{4\pi}{3k_\theta}(0,1),\quad \bm{G}^\text{M}_1=\sqrt{3}k_\theta(1,0) ,\quad\bm{G}^\text{M}_2=\sqrt{3}k_\theta(-\frac{1}{2},\frac{\sqrt{3}}{2})
\end{gather}
where $k_\theta=2k_D\sin\frac{\theta}{2}$, and $a$ is the graphene C-C bond length. The intracell coordinates for the $A$ and $B$ atoms of graphene are $\bm{\tau}_A=a(\frac{\sqrt{3}}{2},\frac{1}{2})$ and $\bm{\tau}_B=a(\frac{\sqrt{3}}{2},\frac{1}{2})$. $\bm{K}_D=\frac{4\pi}{3\sqrt{3}a}(1,0)$ is the valley-$K$ (i.e. $\tau=+$) Dirac point for untwisted graphene. The Dirac point in valley $K'$ ($\tau=-$) is found by time reversal.

The intralayer kinetic term in valley $\tau$ is given by the Dirac cones
\begin{gather}
    \bra{\bm{k},l}\hat{H}^\textrm{BM}\ket{\bm{k}',l}=\hbar v_F\bm{k}\cdot
    \begin{pmatrix}
    \tau\sigma_x\\
    -\sigma_y
    \end{pmatrix}\delta_{\bm{k}\bm{k}'}
\end{gather}
where the momenta $\bm{k}$ above are measured from the Dirac points of the respective layer. Note that we have neglected the small Pauli rotation terms which is required to strictly satisfy particle-hole symmetry $\hat{\mathcal{P}}$.

The interlayer hopping term for valley $+$ is
\begin{gather}
\bra{\bm{k},1}\hat{H}^\textrm{BM}\ket{\bm{k}',2} = 
T_1\delta_{\bm{k}-\bm{k}',\bm{0}} + T_2\delta_{\bm{k}-\bm{k}',\bm{G}^\text{M}_1+\bm{G}^\text{M}_2} + T_3\delta_{\bm{k}-\bm{k}',\bm{G}^\text{M}_2}\\
T_1 = \begin{pmatrix}w_{AA}&w_{AB}\\w_{AB}&w_{AA}\end{pmatrix}\\
T_2 = \begin{pmatrix}w_{AA}&w_{AB}e^{i\phi}\\w_{AB}e^{-i\phi}&w_{AA}\end{pmatrix}\\
T_3 = \begin{pmatrix}w_{AA}&w_{AB}e^{-i\phi}\\w_{AB}e^{i\phi}&w_{AA}\end{pmatrix}\\
\phi=\frac{2\pi}{3}
\end{gather}
where the matrices act in sublattice space. The equations in valley $K'$ can be found by time-reversal. Note that now, $\bm{k}$ in both layers is measured with respect to a common basepoint. Therefore the hopping term is moir\'e-periodic.
 
By solving the BM model with a momentum cutoff and projecting to the central bands, one obtains a set of Bloch functions (8 bands in total including spin and valley). The Chern basis is constructing by diagonalizing the sublattice operator $\sigma_z$ projected onto the central subspace. Combining the `sublattice', valley, and spin indices into a common index $\alpha$, the Bloch functions and associated creation operators can be written as $\ket{\psi_\alpha(\bm{k})}=e^{i\bm{k}\cdot\hat{\bm{r}}}\ket{u_{\alpha}(\bm{k})}$ and $\hat{d}^\dagger_{\alpha}(\bm{k})$. The Chern number of the bands are $C=\tau_z\sigma_z$, and the form factors are $[\Lambda_{\bm{q}}(\bm{k})]_{\alpha\beta}=\braket{u_\alpha({\bk})|u_\beta({\bk+\bq})}$ which satisfy $\Lambda_{\bq}(\bk)=\Lambda^\dagger_{-\bq}(\bk+\bq)$.
 
In the Chern basis (with appropriate gauge-fixing, see Ref.~\onlinecite{Bultinck_2020}), the form factors are heavily constrained by $\hat{C}_{2z}\hat{\mathcal{T}}=\sigma_x\hat{\mathcal{K}}$ and (anticommuting) $\hat{\mathcal{P}}\hat{\mathcal{T}}=\tau_y\sigma_y$ symmetries. The sublattice-symmetric and antisymmetric parts of the form factors are 
\begin{equation}
    \Lambda^\text{S}_{\bq }(\bk)=F^\text{S}_{\bq }(\bk)e^{i\phi^\text{S}_{\bq }(\bk)\sigma_z\tau_z},\quad
    \Lambda^\text{A}_{\bq }(\bk)=\sigma_x\tau_zF^\text{A}_{\bq }(\bk)e^{i\phi^\text{A}_{\bq }(\bk)\sigma_z\tau_z}.
\end{equation}
 
\subsection{Strong-coupling form}
Consider first the normal-ordered interaction Hamiltonian in the projected subspace (this will later be augmented by interaction-induced corrections)
\begin{align}
    \hat{H}'&=\hat{H}^\text{BM}+\hat{H}^\text{int,n-o}\\
    &=\sum_{\bk}\hat{\bm{d}}^\dagger({\bk}) h^\text{BM}(\bk)\hat{\bm{d}}({\bk})+\frac{1}{2A}\sum_{\bm{q}}V_{\bm{q}}:\hat{\rho}^\dagger_{\bm{q}}\hat{\rho}_{\bm{q}}:\\
    &=\sum_{\bk}\hat{\bm{d}}^\dagger({\bk}) h^\text{BM}(\bk)\hat{\bm{d}}({\bk})+\frac{1}{2A}\sum_{\bm{q}\bk\bk'\{\alpha\}}V_{\bq}\,[\Lambda_{\bq}(\bk)]_{\alpha\beta}[\Lambda_{-\bq}(\bk')]_{\gamma\delta}\,
    \hat{d}^\dagger_\alpha(\bk)\hat{d}^\dagger_\gamma(\bk')\hat{d}_\delta(\bk'-\bq)\hat{d}_\beta(\bk+\bq)\\
    \hat{\rho}_{\bq}&=\sum_{\bk}\hat{\bd}^\dagger({\bk})\Lambda_{\bq}(\bk)\hat{\bd}({\bk+\bq})
\end{align}
where $h^\text{BM}$ are the BM kinetic matrix elements in the Chern basis. Consider decoupling the normal-ordered interaction term with the projector $P_{\alpha\beta}(\bk)=\langle \hat{d}^\dagger_{\alpha}(\bk)\hat{d}_\beta(\bk) \rangle$ 
\begin{equation}
h^{\text{HF,int}}[P](\bk)=\frac{1}{A}\sum_{\bG}V_{\bG}\Lambda_{\bG}(\bk)\sum_{\bk'}\text{Tr}[P(\bk')\Lambda^*_{\bG}(\bk')]-\frac{1}{A}\sum_{\bq}V_{\bq}\Lambda_{\bq}(\bk)P^T(\bk+\bq)\Lambda_{\bq}^\dagger(\bk).
\end{equation} 
 
To complete the Hamiltonian, we need to account for the interaction contribution of the remote bands, as well as the fact that some interactions have been double-counted (for example the normal-ordered interaction is not particle-hole symmetric about neutrality due to the self direct and exchange interaction of the filled central bands). There is no unique prescription for doing this, but a convenient parameterization is in terms of some reference projector $P^0$. Assuming that the net contributions arising from Bloch states in the remote bands cancel, the total Hamiltonian can be written
\begin{align}
    \hat{H}&=\hat{H}'-\hat{H}^{\text{HF,int}}[P^0]\\
    &=\sum_{\bk}\hat{\bm{d}}^\dagger({\bk}) h(\bk)\hat{\bm{d}}({\bk})+\frac{1}{2A}\sum_{\bm{q}}V_{\bm{q}}:\hat{\rho}^\dagger_{\bm{q}}\hat{\rho}_{\bm{q}}:\\
    h(\bk)&=h^{\text{BM}}(\bk)-h^{\text{HF,int}}[P^0](\bk).
\end{align}
In this way at the reference density $P=P^\text{0}$, the HF spectrum is the same as that of the BM model.

We now recast the Hamiltonian in `strong-coupling' form, which makes the hierarhcy of scales more transparent. Letting $P^0(\bk)=\frac{1}{2}[1+Q^0(\bk)]$, the expression for $h(\bk)$ becomes
\begin{equation}\label{appeq:h}
    h(\bk)=h^{\text{BM}}(\bk)-\frac{1}{2A}\sum_{\bG}V_{\bG}\Lambda_{\bG}(\bk)\sum_{\bk'}\text{Tr}[\Lambda^*_{\bG}(\bk')]+\frac{1}{2A}\sum_{\bq}V_{\bq}\Lambda_{\bq}(\bk)\Lambda_{\bq}^\dagger(\bk)+\frac{1}{2A}\sum_{\bq}V_{\bq}\Lambda_{\bq}(\bk){Q^0}(\bk+\bq)^T\Lambda_{\bq}^\dagger(\bk)
\end{equation}
where a term vanished because $P^0$ is assumed to be a neutral projector arising from a $\hat{C}_{2z}\hat{\mathcal{T}}$ and $\hat{\mathcal{P}}\hat{\mathcal{T}}$ symmetric Hamiltonian. The normal-ordered interaction can be rewritten
\begin{equation}\label{appeq:intno}
    \hat{H}^{\text{int,n-0}}=\frac{1}{2A}\sum_{\bq\bk\bk'}V_{\bq }[\hat{\bd}^\dagger_{\bk }\Lambda_{\bq }(\bk)\hat{\bd}_{\bk+\bq }][\hat{\bd}^\dagger_{\bk' }\Lambda_{-\bq }(\bk')\hat{\bd}_{\bk'-\bq }]
    -\frac{1}{2A}\sum_{\bq\bk}V_{\bq }\hat{\bd}^\dagger_{\bk}\Lambda_{\bq }(\bk)\Lambda_{-\bq }(\bk+\bq )\hat{\bd}_{\bk}.
\end{equation}
The second term of Eq.~\ref{appeq:intno} cancels with the third term of Eq.~\ref{appeq:h}. Collecting the rest of the terms, we obtain the strong-coupling form of the total Hamiltonian
\begin{gather}
    \hat{H}=\sum_{\bk}\hat{\bm{d}}^\dagger({\bk}) h^{\text{SP}}(\bk)\hat{\bm{d}}({\bk})+\frac{1}{2A}\sum_{\bm{q}}V_{\bm{q}}\delta\hat{\rho}^\dagger_{\bm{q}}\delta\hat{\rho}_{\bm{q}}\\
    h^\text{SP}(\bk)=h^\text{BM}(\bk)+\frac{1}{2A}\sum_{\bq}V_{\bq}\Lambda_{\bq}(\bk){Q^0}(\bk+\bq)^T\Lambda_{\bq}^\dagger(\bk)\\
    \delta{\hat{\rho}}_{\bq }=\hat{\rho}_{\bq }-\bar{\rho}_{\bq }, \quad 
    \bar{\rho}_{\bq }=\frac{1}{2}\sum_{\bG\bk}\delta_{\bG,\bq }\text{Tr}\Lambda_{\bG }(\bk).
\end{gather}
 
Dividing the effective single-particle term into inter-sublattice and (sub-dominant) intra-sublattice pieces
\begin{equation}
    h^{\text{SP}}(\bk)=h_0(\bk)\tau_z + h_x(\bk)\sigma_x + h_y(\bk)\sigma_y\tau_z,
\end{equation}
we can define the energy scales of the strong-coupling hierarchy
\begin{align}
    U_\text{S}&=\frac{1}{2AN}\sum_{\bk \bq }V_{\bq }|F^\text{S}_{\bq }(\bk)|^2\\
    U_\text{A}&=\frac{1}{2AN}\sum_{\bk \bq }V_{\bq }\left(|F^\text{S}_{\bq }(\bk)F^\text{A}_{\bq }(\bk)|+|F^\text{A}_{\bq }(\bk)|^2\right)\\
    t_\text{S}&=\frac{1}{N}\sum_{\bk }|h_x(\bk)+ih_y(\bk) |\\
    t_\text{A}&=\frac{1}{N}\sum_{\bk }|h_0(\bk)|.
\end{align}
 
Similarly, the NLSM parameters are
\begin{align}
    E_J&=JA_{\text{UC}}=\frac{1}{N}\sum_{\bk\bk'}[h_x(\bk)+ih_y(\bk)][\mathscr{H}_\text{eh}^{-1}]_{\bk\bk'}[h_x(\bk')+ih_y(\bk')]\\
    \text{where }[\mathscr{H}_\text{eh}]_{\bk\bk'}&=\frac{1}{A}\sum_{\bq}V_{\bq }|F^{\text{S}}_{\bq }(\bk)|^2\left[\delta_{\bk\bk'}-\delta_{\bk',\lfloor \bk+\bq\rfloor}e^{2i\phi^\text{S}_{\bq }(\bk)}\right]\\
    E_\lambda&=\lambda A_{\text{UC}}=\frac{1}{2AN}\sum_{\bk\bq }V_{\bq }|F^\text{A}_{\bq }(\bk)|^2\\
    \rho_{\text{ps}}&=-\frac{1}{8A^2}\nabla_{\bm{q}'}^2\sum_{\bm{k},\bm{q}}V_{\bm{q}}M_{\bm{k}-\bm{q}-\bm{q}'}(\bm{q},\bm{q}')M_{\bm{k}}(-\bm{q},-\bm{q}')\big|_{\bm{q}'=0}\\
    M_{\bm{k}}(\bm{q},\bm{q}')&=\Lambda_{\bm{q}}(\bm{k})\Lambda_{\bm{q}'}(\bm{k}+\bm{q})-\Lambda_{\bm{q}'}(\bm{k})\Lambda_{\bm{q}}(\bm{k}+\bm{q}').
\end{align}
The pseudospin stiffness $\rho_{\text{ps}}$ is a property of the fully symmetric limit. 

\section{Effective skyrmion hopping/hybridization models}\label{app:effective}
In this appendix, we describe the construction of effective hopping/hybridization models for localized skyrmions about the KIVC state at even integer filling. 
 
\subsection{One- and two-body matrix elements}
We first review the general expressions for the matrix elements of one- and two-body operators between different Slater determinants. Consider two general Slater determinants with the same particle number
\begin{gather}
    \ket{A}=\prod_{m\leq N}a^\dagger_m\ket{0},\quad \ket{B}=\prod_{n\leq N}b^\dagger_n\ket{0}\\
    a^\dagger_m=\sum_{\alpha}A_{\alpha m}d^\dagger_\alpha\\
    b^\dagger_n=\sum_{\alpha}B_{\beta n}d^\dagger_\beta
\end{gather}
where the $d^\dagger_\alpha$ operators represent some natural underlying basis of the Hamiltonian. The $m,n$ indices above only run up to $N$, so the transformation matrices $A,B$ are of dimension $N_b\times N$, where $N_b$ is the Hilbert space dimension. An important quantity is the $N\times N$ matrix $S_{mn}$ which relates the natural bases of the two states $\ket{A}$ and $\ket{B}$
\begin{gather}
    \{ a^\dagger_m,b_n\}=(A^T B^*)_{mn}\equiv S^*_{mn}\\
    \{ a_m,b^\dagger_n\}=(A^\dagger B^)_{mn}\equiv S_{mn}\\
    \braket{A|B}=\det S^{A,B},
\end{gather}
where the superscripts in $S^{A,B}$ emphasize that the determinant is taken over the filled spaces of $\ket{A}$ and $\ket{B}$ (i.e. the entire matrix $S$).
 
We now consider the one-body matrix elements $P^{A,B}_{\alpha,\beta}=\bra{A}d^\dagger_\alpha d_\beta\ket{B}$
\begin{align}
    P^{A,B}_{\alpha,\beta}&=\delta_{\alpha\beta}\det S^{A,B}-\bra{A}d_\beta d^\dagger_\alpha\ket{B}\\
    &=\delta_{\alpha\beta}\det S^{A,B}-\det \begin{pmatrix}
    S^{\beta,\alpha} & S^{\beta,B} \\ S^{A,\alpha} & S^{A,B}
    \end{pmatrix}\\
    &=\delta_{\alpha\beta}\det (A^\dagger B)-\det \begin{pmatrix}
    \delta_{\beta,\alpha} & \left[B\right]_{\beta,\cdot} \\ \left[A^\dagger\right]_{\cdot,\alpha} & \left[A^\dagger B\right]_{\cdot,\cdot}\end{pmatrix}\\
    &=\delta_{\alpha\beta}\det (A^\dagger B)-\det (A^\dagger B)\left(\delta_{\alpha\beta}-\left[B(A^\dagger B)^{-1}A^\dagger\right]_{\beta,\alpha}\right)\\
    &=\det (A^\dagger B)\left[B(A^\dagger B)^{-1}A^\dagger\right]_{\beta,\alpha}
\end{align}
where $\cdot$ indicates vector/matrix structure in that index.
 
Now the two-body matrix elements
\begin{align}
    P^{A,B}_{\alpha,\beta;\gamma,\delta}&\equiv \bra{A}d^\dagger_\alpha d^\dagger_\beta d_\gamma d_\delta \ket{B}\\
    &=P^{A,B}_{\alpha,\delta}\delta_{\beta\gamma}+P^{A,B}_{\beta,\gamma}\delta_{\alpha\delta}-P^{A,B}_{\beta,\delta}\delta_{\alpha\gamma}-P^{A,B}_{\alpha,\gamma}\delta_{\beta\delta}+(\delta_{\alpha\gamma}\delta_{\beta\delta}-\delta_{\alpha\delta}\delta_{\beta\gamma})\det(A^\dagger B)+\det \begin{pmatrix}
    S^{(\delta,\gamma),(\alpha,\beta)} & S^{(\delta,\gamma),B} \\
    S^{A,(\alpha,\beta)} & S^{A,B}
    \end{pmatrix}.
\end{align}
But using $S^{(\delta,\gamma),(\alpha,\beta)}=\delta_{\alpha\delta}\delta_{\beta\gamma}-\delta_{\alpha\gamma}\delta_{\beta\delta}$ and an identity for determinants of block matrices
\begin{align}
    \det \begin{pmatrix}
    S^{(\delta,\gamma),(\alpha,\beta)} & S^{(\delta,\gamma),B} \\
    S^{A,(\alpha,\beta)} & S^{A,B}
    \end{pmatrix}&=\bra{A}d_\gamma d_\delta d^\dagger_\alpha d^\dagger_\beta \ket{B}\\
    &=\det(S^{A,B})\det\left(S^{(\delta,\gamma),(\alpha,\beta)}-S^{(\delta,\gamma),B}(S^{A,B})^{-1}S^{A,(\alpha,\beta)}\right)\\
    &=\det(S^{A,B})\det\left(\begin{pmatrix}
    \delta_{\alpha\delta}&\delta_{\beta\delta}\\
    \delta_{\alpha\gamma}&\delta_{\beta\gamma}
    \end{pmatrix}
    -\left[B(A^\dagger B)^{-1}A^\dagger)\right]_{(\delta,\gamma),(\alpha,\beta)}\right).
\end{align}
Hence we obtain the final result
\begin{align}
    P^{A,B}_{\alpha,\beta;\gamma,\delta}=&P^{A,B}_{\alpha,\delta}\delta_{\beta\gamma}+P^{A,B}_{\beta,\gamma}\delta_{\alpha\delta}-P^{A,B}_{\beta,\delta}\delta_{\alpha\gamma}-P^{A,B}_{\alpha,\gamma}\delta_{\beta\delta}+\det(A^\dagger B)(\delta_{\alpha\gamma}\delta_{\beta\delta}-\delta_{\alpha\delta}\delta_{\beta\gamma})\\
    &+\det(A^\dagger B)\det\left(\begin{pmatrix}
    \delta_{\alpha\delta}&\delta_{\beta\delta}\\
    \delta_{\alpha\gamma}&\delta_{\beta\gamma}
    \end{pmatrix}
    -\left[B(A^\dagger B)^{-1}A^\dagger)\right]_{(\delta,\gamma),(\alpha,\beta)}\right).
\end{align}

\subsection{Application to TBG --- single-component hopping}\label{subapp:hopping}
We first discuss the case of a single type of localized skyrmion (e.g. we neglect partners related by point group symmetries in the case of paired skyrmions with substantial ellipticity). Recall how a localized skyrmion $\ket{\phi_{\bm{0}}}$ is constructed from the eigenvectors $v^n_{\bk,\tau,a}(\bm{0})$ of the HF Hamiltonian (the $\bm{0}$ denotes that the skyrmion is centered at some arbitrarily chosen origin moir\'e site)
\begin{gather}
    \sum_{\bk' \tau' b}\mathcal{H}^{\text{HF}}_{\bk \tau a;\bk' \tau' b}[P(\bm{0})]v^n_{\bk',\tau',b}(\bm{0})=\epsilon^n v^n_{\bk,\tau,a}(\bm{0})\\
    \text{i.e. fill orbitals }\sum_{\bk \tau a}v^n_{\bk,\tau,a}(\bm{0})d^\dagger_{\bk \tau a}\text{ up to some $n$}.
\end{gather}
We construct skyrmions $\ket{\phi_{\bm{R}}}$ at all other moir\'e sites $\bm{R}$ by boosting the eigenvectors $v^n_{\bk,\tau,a}(\bm{R})=e^{i\bk \bm{R}}v^n_{\bk,\tau,a}(\bm{0})$. Treating the set of eigenvectors as a matrix $\left[v(\bm{R})\right]_{\alpha,n}=v_{\alpha}^n(\bm{R})$, where $\alpha=(\bk,\tau,a)$, we obtain expressions for the overlap, one-body, and two-body correlators
\begin{align}
    S(\bm{R})&=\braket{\phi_{\bm{0}}|\phi_{\bm{R}}}=\det \left[v^\dagger(\bm{0}) v(\bm{R})\right]\\
    P^1_{\alpha,\beta}(\bm{R})&=\bra{\phi_{\bm{0}}}d^\dagger_\alpha d_\beta \ket{\phi_{\bm{R}}}\\
    &=S(\bm{R})\left[v(\bm{R})\left(v(\bm{0})^\dagger v(\bm{R})\right)^{-1}v(\bm{0})^\dagger\right]_{\beta,\alpha}\\
    P^2_{(\alpha,\beta);(\gamma,\delta)}(\bm{R})&=\bra{\phi_{\bm{0}}}d^\dagger_\alpha d^\dagger_\beta d_\gamma d_\delta \ket{\phi_{\bm{R}}}\\
    &=S(\bm{R})(\delta_{\alpha\gamma}\delta_{\beta\delta}-\delta_{\alpha\delta}\delta_{\beta\gamma})+P^{1}_{\alpha,\delta}(\bm{R})\delta_{\beta\gamma}+P^{1}_{\beta,\gamma}(\bm{R})\delta_{\alpha\delta}-P^{1}_{\beta,\delta}(\bm{R})\delta_{\alpha\gamma}-P^{1}_{\alpha,\gamma}(\bm{R})\delta_{\beta\delta}\\
    &\phantom{=}\,\,\,+\det(S(\bm{R}))\det\left(\begin{pmatrix}
    \delta_{\alpha\delta}&\delta_{\beta\delta}\\
    \delta_{\alpha\gamma}&\delta_{\beta\gamma}
    \end{pmatrix}
    -\left[v(\bm{R})(v(\bm{0})^\dagger v(\bm{R}))^{-1}v(\bm{0})^\dagger)\right]_{(\delta,\gamma),(\alpha,\beta)}\right).
\end{align}
The other correlators (say between a skyrmion at $\bm{R}_1$ and $\bm{R}_2$ can be obtained by translation invariance of the Hamiltonian.

The hopping model is constructed using the overlaps and matrix elements of the localized skyrmions
\begin{gather}
    S_{\bm{R},\bm{R}'}=\braket{\phi_{\bm{R}}|\phi_{\bm{R}'}}=S(\bm{R}-\bm{R}')\\
    H_{\bm{R},\bm{R}'}=\bra{\phi_{\bm{R}}}\hat{H}\ket{\phi_{\bm{R}'}}=H(\bm{R}-\bm{R}')
\end{gather}
where the Hamiltonian matrix elements are
\begin{align}
    H(\bm{R})&=\sum_{\alpha\beta}H^{\text{SP}}_{\alpha,\beta}P^1_{\alpha,\beta}(\bm{R})\\
    &+\frac{1}{2A}\sum_{\tau\tau'abcd}\sum_{\bm{k}^\alpha\bm{k}^\beta\bm{q}\bm{G}}V(\bm{q},\bm{G})\lambda_{\tau,ab}(\bm{k}^\alpha;\bm{q},\bm{G})\lambda^*_{\tau',dc}(\bm{k}^\beta;\bm{q},\bm{G})P^2_{(\alpha,\beta);(\gamma,\delta)}(\bm{R})\\
    \alpha&=(\bm{k}^\alpha,\tau,a)\\
    \beta&=(\bm{k}^\beta+\bm{q},\tau',c)\\
    \gamma&=(\bm{k}^\beta,\tau',d)\\
    \delta&=(\bm{k}^\alpha+\bm{q},\tau,b)
\end{align}
and $H^\text{SP}$ is the effective one-body term which includes interaction-induced dispersion corrections.

The resulting model is solved by simple diagonalization
\begin{gather}
    \ket{\psi}=\sum_{\bm{R}}c_{\bm{R}}\ket{\phi_{\bm{R}}}\\
    \sum_{\bm{R}'}[S^{-1}H]_{\bm{R},\bm{R}'}c_{\bm{R}'}=Ec_{\bm{R}}.
\end{gather}
Due to moir\'e translation invariance, this leads to `Bloch skyrmions' with plane wave coefficients and a skyrmion bandstructure in the mBZ.

\subsection{Application to TBG --- multi-component hybridization}
Next we discuss the case of multiple components of localized skyrmion, for example the $\hat{C}_3$-related partners of a given paired skyrmion. Here we ignore the hopping problem, and instead consider how the components (all localized at the same moir\'e site) hybridize. We consider $N_c$ components indexed by $l$, with corresponding skyrmion Slater determinants $\ket{\phi_l}$ and HF eigenvectors $v^n_{\bm{k},\tau,a}(l)$. Treating the set of eigenvectors as a matrix $\left[v(l)\right]_{\alpha,n}=v_{\alpha}^n(l)$, we obtain expressions for the overlap, one-body, and two-body correlators
\begin{align}
    S(l,l')&=\braket{\phi_{l}|\phi_{l'}}=\det \left[v^\dagger(l) v(l')\right]\\
    P^1_{\alpha,\beta}(l,l')&=\bra{\phi_{l}}d^\dagger_\alpha d_\beta \ket{\phi_{l'}}\\
    &=S(l,l')\left[v(l')\left(v(l)^\dagger v(l')\right)^{-1}v(l)^\dagger\right]_{\beta,\alpha}\\
    P^2_{(\alpha,\beta);(\gamma,\delta)}(l,l')&=\bra{\phi_{l}}d^\dagger_\alpha d^\dagger_\beta d_\gamma d_\delta \ket{\phi_{l'}}\\
    &=S(l,l')(\delta_{\alpha\gamma}\delta_{\beta\delta}-\delta_{\alpha\delta}\delta_{\beta\gamma})+P^{1}_{\alpha,\delta}(l,l')\delta_{\beta\gamma}+P^{1}_{\beta,\gamma}(l,l')\delta_{\alpha\delta}-P^{1}_{\beta,\delta}(l,l')\delta_{\alpha\gamma}-P^{1}_{\alpha,\gamma}(l,l')\delta_{\beta\delta}\\
    &\phantom{=}\,\,\,+\det(S(l,l'))\det\left(\begin{pmatrix}
    \delta_{\alpha\delta}&\delta_{\beta\delta}\\
    \delta_{\alpha\gamma}&\delta_{\beta\gamma}
    \end{pmatrix}
    -\left[v(l')(v(l)^\dagger v(l'))^{-1}v(l)^\dagger)\right]_{(\delta,\gamma),(\alpha,\beta)}\right).
\end{align}
The Hamiltonian is constructed in an analogous way to App.~\ref{subapp:hopping}.

\subsection{Application to TBG --- multi-component hopping}
The most general case if we consider multiple components and include hopping. We consider skyrmions with position $\bm{R}$ with $N_c$ components indexed by $l$. We have corresponding skyrmion Slater determinants $\ket{\phi_{\bm{R};l}}$ and HF eigenvectors $v^n_{\bm{k},\tau,a}(\bm{R};l)$. We obtain expressions for the overlap, one-body, and two-body correlators
\begin{align}
    S(\bm{R};l,l')&=\braket{\phi_{\bm{0};l}|\phi_{\bm{R};l'}}=\det \left[v^\dagger(\bm{0};l) v(\bm{R};l')\right]\\
    P^1_{\alpha,\beta}(\bm{R};l,l')&=\bra{\phi_{\bm{0};l}}d^\dagger_\alpha d_\beta \ket{\phi_{\bm{R};l'}}\\
    &=S(\bm{R};l,l')\left[v(\bm{R};l')\left(v(\bm{0};l)^\dagger v(\bm{R};l')\right)^{-1}v(\bm{0};l)^\dagger\right]_{\beta,\alpha}\\
    P^2_{(\alpha,\beta);(\gamma,\delta)}(\bm{R};l,l')&=\bra{\phi_{\bm{0};l}}d^\dagger_\alpha d^\dagger_\beta d_\gamma d_\delta \ket{\phi_{\bm{R};l'}}\\
    &=S(\bm{R};l,l')(\delta_{\alpha\gamma}\delta_{\beta\delta}-\delta_{\alpha\delta}\delta_{\beta\gamma})\\
    &\phantom{=}\,\,\,+P^{1}_{\alpha,\delta}(\bm{R};l,l')\delta_{\beta\gamma}+P^{1}_{\beta,\gamma}(\bm{R};l,l')\delta_{\alpha\delta}-P^{1}_{\beta,\delta}(\bm{R};l,l')\delta_{\alpha\gamma}-P^{1}_{\alpha,\gamma}(\bm{R};l,l')\delta_{\beta\delta}\\
    &\phantom{=}\,\,\,+\det(S(\bm{R};l,l'))\det\left(\begin{pmatrix}
    \delta_{\alpha\delta}&\delta_{\beta\delta}\\
    \delta_{\alpha\gamma}&\delta_{\beta\gamma}
    \end{pmatrix}
    -\left[v(\bm{R};l')(v(\bm{0};l)^\dagger v(\bm{R};l'))^{-1}v(\bm{0};l)^\dagger)\right]_{(\delta,\gamma),(\alpha,\beta)}\right).
\end{align}
The effective model can be easily solved with a Bloch ansatz
\begin{gather}
    \ket{\psi_{\bm{k}}}=\frac{1}{\sqrt{N_{\text{UC}}}}\sum_{l\bm{R}}f_le^{i\bm{k}\bm{R}}\ket{\phi_{\bm{R};l}}\\
    \sum_{l'}\left[\tilde{S}^{-1}(\bm{k})\tilde{H}(\bm{k})\right]_{l,l'}f_{l'}=Ef_{l}\\
    \tilde{S}_{l,l'}(\bm{k})=\sum_{\bm{R}}S_{l,l'}(\bm{R})e^{-i\bm{k}\bm{R}}\\
    \tilde{H}_{l,l'}(\bm{k})=\sum_{\bm{R}}H_{l,l'}(\bm{R})e^{-i\bm{k}\bm{R}}
\end{gather}

 \end{appendix}
 
\end{widetext} 
\end{document}